\documentclass[useAMS,usenatbib]{mn2e}
\usepackage{graphicx}

\def\aj{AJ}%
\def\araa{ARA\&A}%
\def\apj{ApJ}%
\def\apjl{ApJ}%
\def\apjs{ApJS}%
\def\aap{A\&A}%
\def\icarus{Icarus}%
\def\mnras{MNRAS}%
\def\nat{Nature}%
\newcommand{\msun}{M_\odot}
\newcommand{\mjup}{M$_{\rm J}$}
\newcommand{\halfmass}{r_{\rm hm}}
\newcommand{\mcluster}{M_{\rm cl}}
\newcommand{\kms}{km\,s$^{-1}$}

\newcommand{\nbps}{N_{\rm BPS}}
\newcommand{\neps}{N_{\rm EPS}}
\newcommand{\nbfp}{N_{\rm BFP}}
\newcommand{\nefp}{N_{\rm EFP}} 

\newcommand{\fbps}{f_{\rm BPS}}
\newcommand{\feps}{f_{\rm EPS}}
\newcommand{\fbfp}{f_{\rm BFP}}
\newcommand{\fefp}{f_{\rm EFP}} 

\newcommand{\fplanetbound}{f_{\rm p,b}}
\newcommand{\fplanetunbound}{f_{\rm p,e}}
\newcommand{\fffbound}{f_{\rm ff,b}}
\newcommand{\fffunbound}{f_{\rm ff,e}}
\newcommand{\nbs}{S_b}
\newcommand{\nes}{S_e}

\newcommand{\nobj}{N_{\rm sp}}
\newcommand{\nstar}{N_{\rm s}}
\newcommand{\nplanet}{N_{\rm p}}

\newcommand{\rhoc}{\rho}

\newcommand{\tdyn}{t_{\rm dyn}}
\newcommand{\trelax}{t_{\rm rlx}}

\title[Planetary systems in young star
clusters]{The dynamical fate of planetary systems in young star
  clusters} \author[Zheng, Kouwenhoven \& Wang]{Xiaochen Zheng$^{1,2}$\thanks{E-mail:
    x.c.zheng1989@gmail.com}, M.B.N. Kouwenhoven$^{1}$ and Long Wang$^{1,2}$\\
  $^1$Kavli Institute for Astronomy and Astrophysics, Peking University, Yiheyuan Lu 5, Haidian District, Beijing 100871, P.R. China \\
  $^2$Department of Astronomy, Peking University, Yiheyuan Lu 5, Haidian District, Beijing 100871, P.R. China}

\begin{document}

\date{}


\maketitle

\label{firstpage}


\begin{abstract}
  We carry out $N$-body simulations to examine the effects of
  dynamical interactions on planetary systems in young open star
  clusters. We explore how the planetary populations in these star clusters evolve, and how this evolution depends on the initial amount of substructure, the virial ratio, the cluster mass and density, and the initial semi-major axis of the planetary systems. The fraction of planetary systems that remains intact as a cluster member, $\fbps$, is generally well-described by the functional form $\fbps= f_{0} (1+[a/a_0]^c)^{-1}$, where $(1-f_0)$ is the fraction of stars that escapes from the cluster, $a_0$ the critical semi-major axis for survival, and $c$ a measure for the width of the transition region. The effect of the initial amount of substructure over time $t$ can be quantified as $\fbps=A(t)+B(D)$, where $A(t)$ decreases nearly linearly with time, and $B(D)$ decreases when the clusters are initially more substructured. Provided that the orbital separation of planetary systems is smaller than the critical value $a_0$, those in clusters with a higher initial stellar density (but identical mass) have a larger probability of escaping the cluster intact. These results help us to obtain a better understanding of the difference between the observed  fractions of exoplanets-hosting stars in star clusters and in the Galactic field. It also allows us to make predictions about the free-floating planet population over time in different stellar environments.
\end{abstract}

\begin{keywords}
  Open clusters and associations: general -- $N$-body simulations -- extra-solar planets -- planetary systems
\end{keywords}


\section{Introduction}

During the last decade, the Kepler space mission and other transit, radial velocity surveys, microlensing, and imaging surveys, resulted in the discovery of thousands of exoplanets and exoplanet candidates \citep[e.g.,][]{borucki2011, batalha2013}, demonstrating that exoplanets in the Galactic field are common. Combined with new developments in computational astrophysics, the prospects for deepening our understanding of the formation and dynamical evolution of exoplanets are promising.

In contrast to the large number of exoplanets found in the Solar neighborhood, only a handfull of exoplanets have been detected in open clusters, such as Hyades \citep{guenther2005}, M37 \citep{hartman2009}, NGC2158 \citep{mochejska2006}, NGC7789 \citep{bramich2006}, NGC1245 \citep{burke2006} and NGC7068 \citep{rosvick2006}, and in globular clusters, such as $\omega$~Centauri \citep{weldrake2008}, 47~Tucanae \citep{weldrake2005}, NGC6378 \citep{nascimbeni2012} and NGC6791 \citep{mochejska2005, montalto2007}. Although the study of \cite{meibom2013} suggests that the planet frequency in open clusters is similar to that in the field, statistically stronger observational evidence is necessary strengthen this argument.

Observations have indicated that most stars in the Galactic field were formed in hierarchically structured star-forming regions \citep{bressert2012}, often (but not always) in embedded clusters \citep[e.g.,][]{clarke2000, lada2003} that may or may not be gravitationally bound \citep[e.g.,][]{kruijssen2012}. This may also be true for our own Sun, which is believed to have been born in a young star cluster with several thousands of stellar
members \citep[see][for discussions]{adams2001, portegies2009, adams2010, dukes2012, pfalzner2013, parker2014}. In such dense stellar environments, close encounters between planetary systems and the stellar populations surrounding them can result in orbital reconfiguration or decay of planetary systems \citep[e.g.,][]{spurzem2009, boley2012, hao2013} and in the subsequent production of free-floating planets. Although free-floating planets are expected to be common in young star clusters and in the Galactic field \citep[e.g.,][]{pacucci2013}, only few have been detected through microlensing campaigns \citep[e.g.,][]{sumi2011, distefano2012, gaudi2012}, and deep imaging surveys of young star clusters \citep[e.g.,][]{lucas2006, quanz2010, delorme2012}. 

The currently known planetary systems and free-floating planets are unlikely to have formed at the places where they are found today, and the orbital properties of these planetary systems may have been affected by their internal dynamical evolution and by gravitational interactions with the environment in which they were born. To understand the
observed orbital properties of exoplanets, it is necessary to consider
the dynamical evolution these systems have undergone since their formation, prior to, during, and after the dynamical process that led to the escape of these planetary systems from their natal environment.

Computer simulations are crucial for developing a good theoretical
understanding of the formation and evolution of planetary systems. Many studies \citep[e.g.,][]{laughlin1998, bonnell2001, smith2001, davies2001, adams2006, fregeau2006, liu2013, hao2013} have focused on the different aspects of planetary dynamics in star clusters. 
In an extensive theoretical study, \cite{spurzem2009} derive collisional cross-sections for changes in the orbital elements during encounters between stars and star-planet systems in star clusters.
Previous studies employing direct $N$-body simulations of single-planet systems in clustered environment have mostly focused on open and globular clusters that are initially in virial equilibrium and have smooth initial density
profile \citep[e.g.,][]{hurley2002,spurzem2009}. Observations, however, indicate that star clusters are preferably born with a
high degree of substructure \citep[e.g.,][]{cartwright2004, schmeja2011} and
often in a subvirial state \citep[e.g.,][]{peretto2006, proszkow2009}.
Inspired by the earlier works of \cite{goodwin2006}, \cite{parkerquanz}, \cite{craig2013} and \cite{pacucci2013}, we carry out $N$-body simulations of clusters with different initial morphologies and initial virial states, and a population of planetary systems with a wide range of semi-major axes.

This article presents a theoretical study on the evolution of planetary populations in star clusters, with a focus on how the dynamical evolution of these planets is affected by the initial properties of the star cluster and the initial semi-major axes of their orbits, and the subsequent dynamical evolution of the liberated population of free-floating planets. 
In Section~\ref{section:method} we describe our methods and assumptions. In Section~\ref{section:results} we present the results of our simulations, and finally, we summarise and discuss findings in Section~\ref{section:conclusions}.



\section{Method and terminology} \label{section:method}


\subsection{Initial conditions}

\begin{table}
  \caption{Initial conditions for our reference model. \label{table:initialconditions} }
 \begin{tabular}{ll}
    \hline 
    Parameter              &   Default value   \\  
    \hline
    Number of stars        & $\nstar = 1000$  \\
    Number of planets      & $\nplanet = 1000$  \\
    Structure              & $D = $1.6, 2.3, 3.0 \\
    Virial state           & $Q = $0.3, 0.5, 0.7  \\
    Stellar mass function  & \cite{kroupa2001}; $0.08\msun \leq M \leq 10\msun$ \\
    Half-mass radius       & 1~pc  \\
    Planet mass            & $M_p=1$~\mjup  \\
    Semi-major axis        & $a=100$~AU \\
    Eccentricity           & $e=0$  \\
    Orbital orientation    & Random  \\
    Orbital phase          & Random  \\
    Simulation time        & $t = 50$~Myr \\
    \hline 
  \end{tabular}
\end{table}

We use the
\texttt{NBODY6} package \citep{aarseth2003} to carry out $N$-body
simulations of evolving star clusters that initially contain a large population of (single-planet) planetary systems. Our initial conditions represent the state of the system shortly after most of the gas is expelled. The initial conditions are generated using the open-source package \texttt{MCluster} \citep{kupper2011}, which we
modified to suit our requirements. Table~\ref{table:initialconditions} provides a summary of the initial conditions for our reference models. Our study was inspired by the work of \cite{parkerquanz}, and our initial conditions resemble theirs. In this paper, however, we study much broader range in planetary orbital separations. Unlike \cite{parkerquanz}, we include an external tidal field, we normalise the initial size of the clusters to their half-mass radii (see the discussion below), and we do not include primordial stellar binaries. All results described in our study represent the averages of twenty realisations of each model, unless stated otherwise. 

Our reference models are open cluster-sized systems composed of $\nobj
= \nstar + \nplanet = 2000$ individual bodies, among which $\nstar=1000$ stars and $\nplanet=1000$ planets, each orbiting a star. 
Stellar masses are drawn from the \cite{kroupa2001} initial mass
function in the range $0.08-10~\msun$. The lower limit
corresponds to the hydrogen-burning limit
\citep[e.g.,][]{karttunen2003}. Our choice for the upper limit is roughly consistent with \cite{weidner2004} and \cite{weidner2013}, although our random sampling of the masses from the IMF results in upper mass limits of $8.6\pm 1.1\msun$ for the individual clusters \citep[and it should be noted that the relation between the stellar upper mass limit and the cluster mass is still under debate, see, e.g.,][]{cervino2013a, cervino2013b}. 
Each planet is assigned a mass of 1~\mjup{} ($\sim 9.5 \times 10^{-4}\msun$). For the purpose of our study, the planets can be considered as test masses, as their effect on the dynamics of the stellar population is negligible. 
The initial total mass of the clusters in our reference model is $\mcluster \approx 600~\msun$, a typical value for open clusters.

Although the vast majority of know exoplanets orbit their host star at much smaller separations, wide planetary-mass objects have been observed \citep[e.g.,][]{lafreniere2008, marois2008, marois2010, lagrange2009, kraus2014, naud2014, bailey2014}, and their existence is also supported by numerical studies involving formation at wide orbits or dynamical interactions in multiple star systems and planetary systems that lead to outward scattering \citep[e.g.,][and references therein]{portegies2005, stamatellos2008, malmberg2011, boss2011, vorobyov2013, li2015}. All stars in our reference model are assigned a planet with semi-major axis $a=100$~AU.
Much larger values of $a$ result in frequent disruptions, while single-planet systems which much smaller values of $a$ are mostly dynamically inert in the stellar environments we model. For a similar study involving planets with $a=5$~AU and $a=30$~AU we refer to the work of \cite{parkerquanz}. Note that among planets with identical $a$, those orbiting more massive stars have a larger collisional cross section (due to gravitational scattering), but also have a larger binding energy that is proportional to the host star's mass. To study the relationship between the stability of planetary systems and the initial semi-major axis, $a$, we also study systems with a wide range of semi-major axes ($1~{\rm AU} \leq a \leq 10\,000~{\rm AU}$) in Section~\ref{section:dependencesma}. We merely use these initial conditions for the semi-major axis to study the dynamical evolution of such systems; we do not imply that
they formed at these semi-major axes. Each planet is assigned a random orbital phase in a circular ($e=0$), randomly oriented orbit.

Using \texttt{MCluster} we generate star clusters
with with varying degrees of substructure, quantified through the fractal (substructure) parameter $D$ \citep[see][for details]{cartwright2004, goodwin2004, allison2009}. We adopt three values, $D = 1.6$ (highly clumpy), $D = 2.3$ (intermediate substructure), and $D = 3.0$
(homogeneous on large scales). The procedure for generating these substructured star clusters in \texttt{MCluster} is fully described in the appendix of \cite{kupper2011}, and briefly summarised below. Initially, a single star is placed at the origin of a box. The box is split up into eight smaller boxes, and in some of these a new star is placed; the probability that a star is placed in a box is $2^{(D-3)}$, where $D$ is the fractal parameter. The boxes containing stars are again subdivided into eight boxes, and the procedure is repeated. During each step, velocities are drawn from a normal distribution centred on the parent body, and subsequently rescaled, so that the group of stars in the eight boxes are in virial equilibrium. After generation of a large number of stars, a subset of $\nstar$ stars within the unit sphere is drawn from the cloud of stars. Finally, the cluster positions are scaled to the desired radius of the cluster, and the velocities are scaled to obtain the desired initial virial ratio. 

The virial ratio $Q$ of a star cluster is defined as $Q = -K/P$, where $K$ and $P$ are the total kinetic energy and potential energy, respectively. We study clusters that are initially in virial equilibrium ($Q = 0.5$), clusters which start in cool collapse ($Q = 0.3$) and clusters that are initially expanding ($Q = 0.7$).  Each cluster is scaled to an initial (three-dimensional) half-mass radius $\halfmass = 1$~pc, which is typical for young star clusters in this mass range \citep[e.g.,][]{lada2003}. Consequently, the radius of the sphere enclosing all stars equals $2^{1/3}$~pc $\approx 1.26$~pc. 
The star clusters are assigned Solar orbits around the Milky Way centre. The tidal radius $R_t$ of a star cluster at any time is then given by
\begin{equation} \label{eq:tidalradius}
	R_t = D_G \left( \frac{\mcluster}{3M_G} \right)^{1/3} 
	\approx 6.65 \left( \frac{\mcluster}{1000~\msun} \right)^{1/3}{\rm pc}
\end{equation}
\citep[e.g.,][]{binneytremaine}. Here we have adopted a Galactocentric distance $D_G \approx 8$~kpc and a Milky Way-like galaxy with an enclosed mass of $M_G=5.8\times10^{11}\msun$. The initial tidal radius is $R_t\approx 5.5$~pc$~\approx 7\halfmass$ for the star clusters modelled in our study. 

All models are evolved for $t=50$~Myr, the time beyond which planetary systems experience little further evolution. For a set of star clusters with an identical number stars, average stellar mass, and virial radius, the $N$-body time is proportional to the physical time scale. It should be noted, however, that for our choice of initial conditions, the virial radius depends on $D$. A star cluster that is homogeneous on large scales $D=3$ has typically a larger virial radius than a highly substructured cluster ($D=1.6$). Consequently, star clusters with smaller $D$ tend to evolve faster. Numerical tests indicate that for our initial conditions the relation between physical time $t_{\rm Myr}$ and $N$-body time $t_{\rm NB}$ can be expressed as 
\begin{equation}
t_{\rm Myr} \approx t_{\rm NB} ( 0.37 + 0.13D )  \ ,
\end{equation}
such that our total integration time of 50~Myr corresponds to roughly 93~$N$-body units for $D=1.6$ and about 66~$N$-body units for $D=3.0$. It should be noted, however, that this relation is approximate due to stochastic variations resulting from the relatively small number of stars in our simulations.


\subsection{Dynamical status of the planets}

\begin{table}
  \caption{Classification of the planets in our simulations. \label{table:classification} }
  \begin{tabular}{lll}
    \hline \hline
     & Orbiting a star & Free-floating \\
    \hline
    Cluster        & Bound planetary              & Bound free-floating   \\
     member        & system (BPS) &                 planet (BFP) \\
    \hline
    Escaped        & Escaped planetary              & Escaped free-floating     \\
    from cluster   & system (EPS) &                   planet (EFP) \\
    \hline \hline
  \end{tabular}
\end{table}

\begin{figure*}
\includegraphics[width=0.8\textwidth]{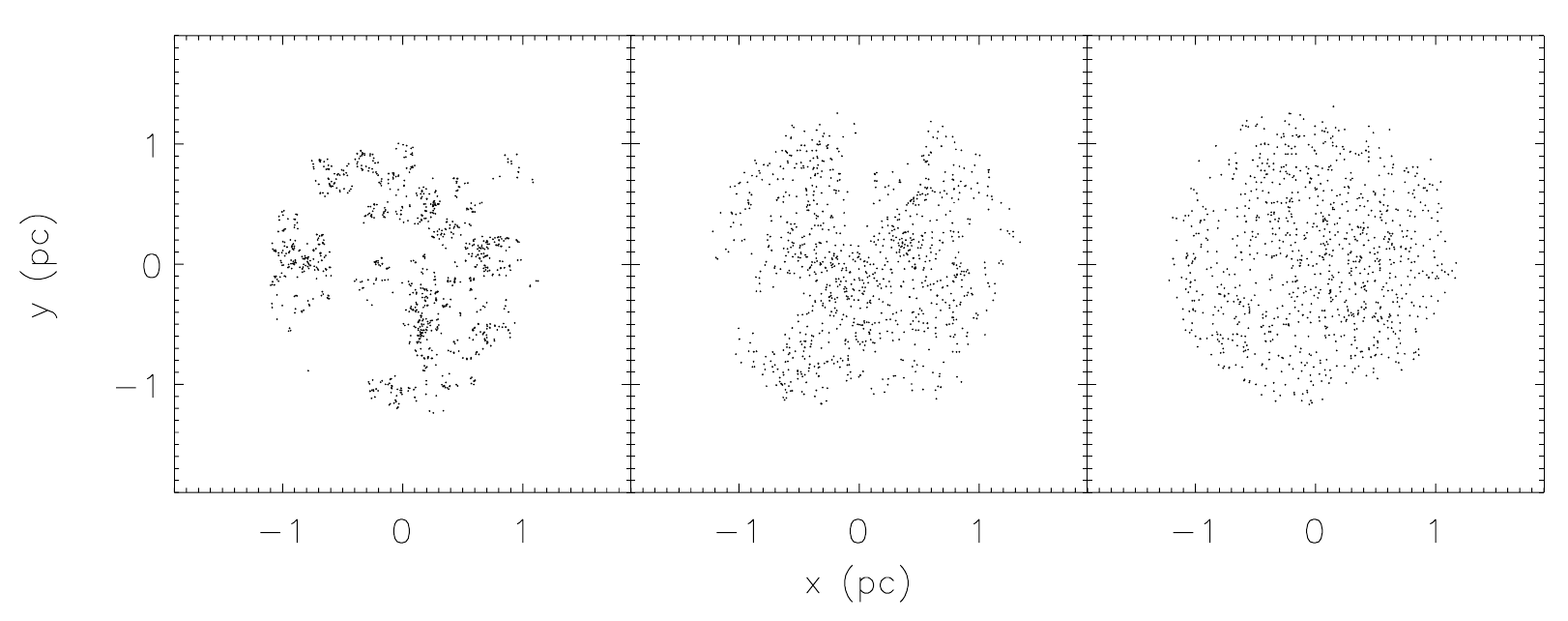}
\caption{Three realisations of the star clusters at $t=0$. The different panels show the position of the stars in the cluster for fractal dimensions (from left to right) $D = 1.6$, $2.3$ and $3.0$, respectively. }
\label{fig:cluster0}
\end{figure*}

In our analysis of the evolution of the planetary population we distinguish between four dynamical categories, based on whether or not a planet is gravitationally bound to a star, and whether or not it is a member of the star cluster. A planet is considered bound to a
star when (i) the star and a planet are each other's mutual nearest
neighbours, and (ii) the gravitational binding energy of the star-planet pair is negative.
Any object (e.g., a star, a binary system, a planetary system, or a
flee-floating planet) is considered as having escaped from the star cluster when each of the three following conditions are satisfied: (i) its velocity is
larger than the star cluster's escape velocity at the location of the
object; (ii) the velocity vector points away from the star
cluster centre, and (iii) the object is located at a distance $r>2R_t$
from the cluster centre, where $R_t$ is the cluster's tidal
radius (Eq.~\ref{eq:tidalradius}). All other objects are considered as gravitationally bound to the star cluster. The four categories and their abbreviations (BPS, BFP, EPS, and EFP) are listed in Table~\ref{table:classification}. None of the planets in our models experiences a physical collision in our simulations, and therefore the total number of planets is conserved at any time:
\begin{equation}
  \nbps + \neps + \nbfp + \nefp = N_{\rm planet} \ .
\end{equation}
We define the corresponding fractions $\fbps$, $\feps$,
$\fbfp$, and $\fefp$ as the ratios between the number of planets in
each category, relative to $N_{\rm planet}$. Since each star is
initially assigned a planetary companion and is initially bound to the star
cluster, the initial fractions are $\fbps=100\%$ and $\feps = \fbfp =
\fefp = 0\%$. After complete dissolution of a star cluster, $\fbps=\fbfp=0\%$ and $\feps+\fefp=100\%$.

A BPS can escape from the star cluster intact as an EPS, or it can be ionised in the star cluster after a close encounter with another star and become a BFP. These BFPs gradually escape from the star cluster and become EFPs \citep{wang2015a}. Capture of single stars into binaries \citep[e.g.,][]{goodman1993, kouwenhoven2010, moeckel2010, moeckel2011}, re-capture of BFPs \citep[e.g.,][]{parkerquanz, perets2012} by stars, and exchange of planets between stars \citep[e.g.,][]{jilkova2015} in the star cluster occur, but these processes are rare. Given enough time, the vast majority of the planets therefore follows one of these  dynamical tracks:
\begin{equation} \label{eq:tracks}
	\begin{tabular}{ll}
		BPS $\rightarrow$ EPS                   & (track 1) \\
		BPS $\rightarrow$ BFP $\rightarrow$ EFP & (track 2) \\
	\end{tabular}
\end{equation}
As a result, $d\nbps/dt\leq 0$, $d\neps/dt\geq 0$, and $d\nefp/dt\geq 0$ at any time. 
In the thousands of simulations we carry out, dynamical binary star systems form, and several of these may host planets in (primordial) S-type configurations or (captured) P-type configurations. In our analysis, we consider these objects as BPSs or EPSs, depending on whether or not they are a cluster member. Although these are interesting processes that warrant further study, they barely affect the analysis of the planetary populations as a whole, and we will therefore not discuss these in this paper.



\section{Results} \label{section:results}

In this section we describe how the planetary population evolves
different environments and how these results depend on the orbital parameters of the planets. The global evolution of the star clusters is described in Section~\ref{section:starclusterevolution}. In Section~\ref{section:dependencetime} and Section~\ref{section:dependencesma} we describe how the planetary population evolves over time, and how this evolution depends on the initial semi-major axis of the planets. Subsequently, we study the dependence on initial amount of substructure in the cluster (Section~\ref{section:dependenced}), and on
the mass of the star clusters (Section~\ref{section:dependenceclustermass}). 


\subsection{Star cluster evolution} \label{section:starclusterevolution}

\begin{figure}
  \begin{tabular}{p{0.2\textwidth}p{0.2\textwidth}}
    \includegraphics[width=0.25\textwidth]{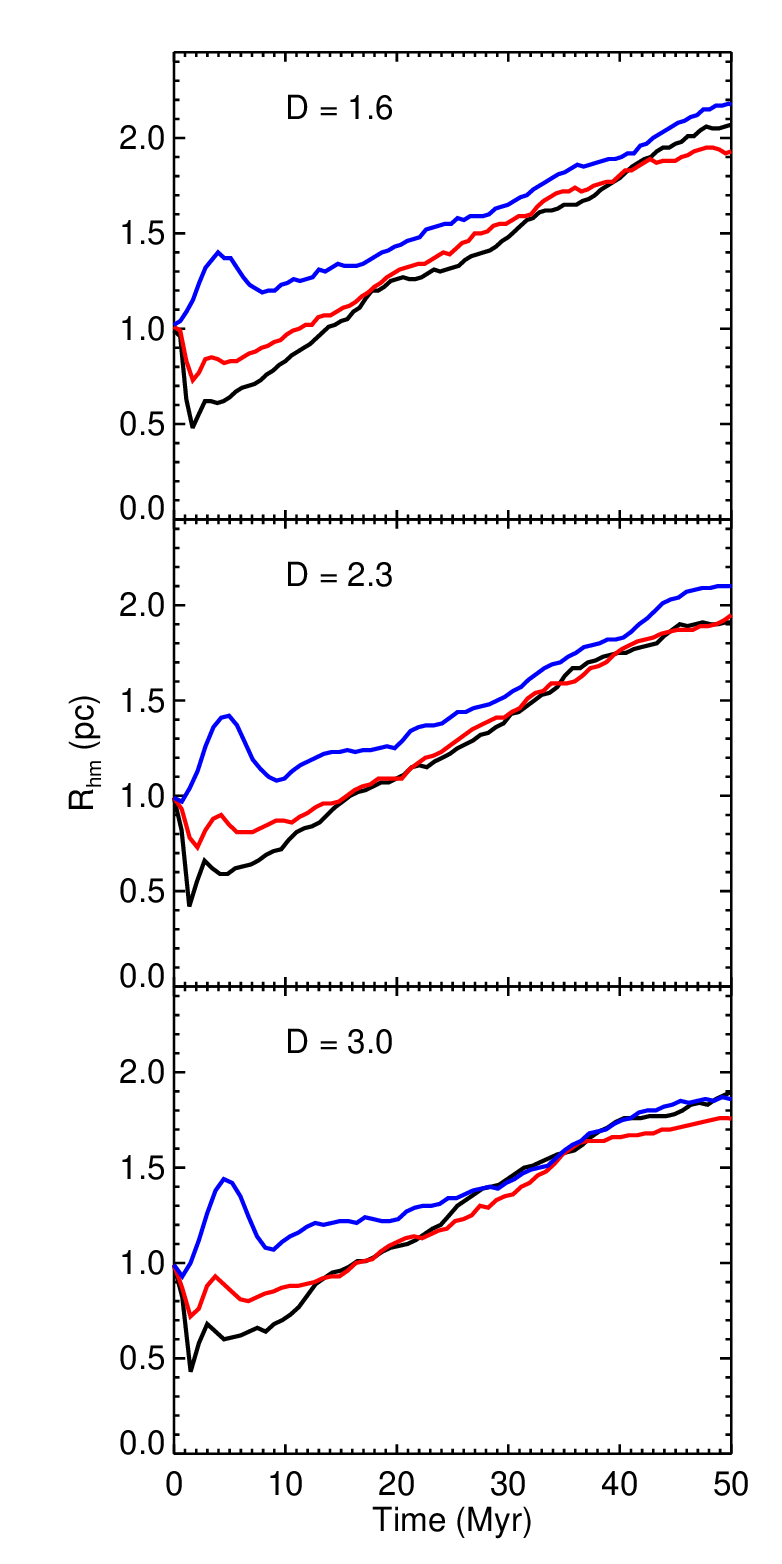} &
    \includegraphics[width=0.25\textwidth]{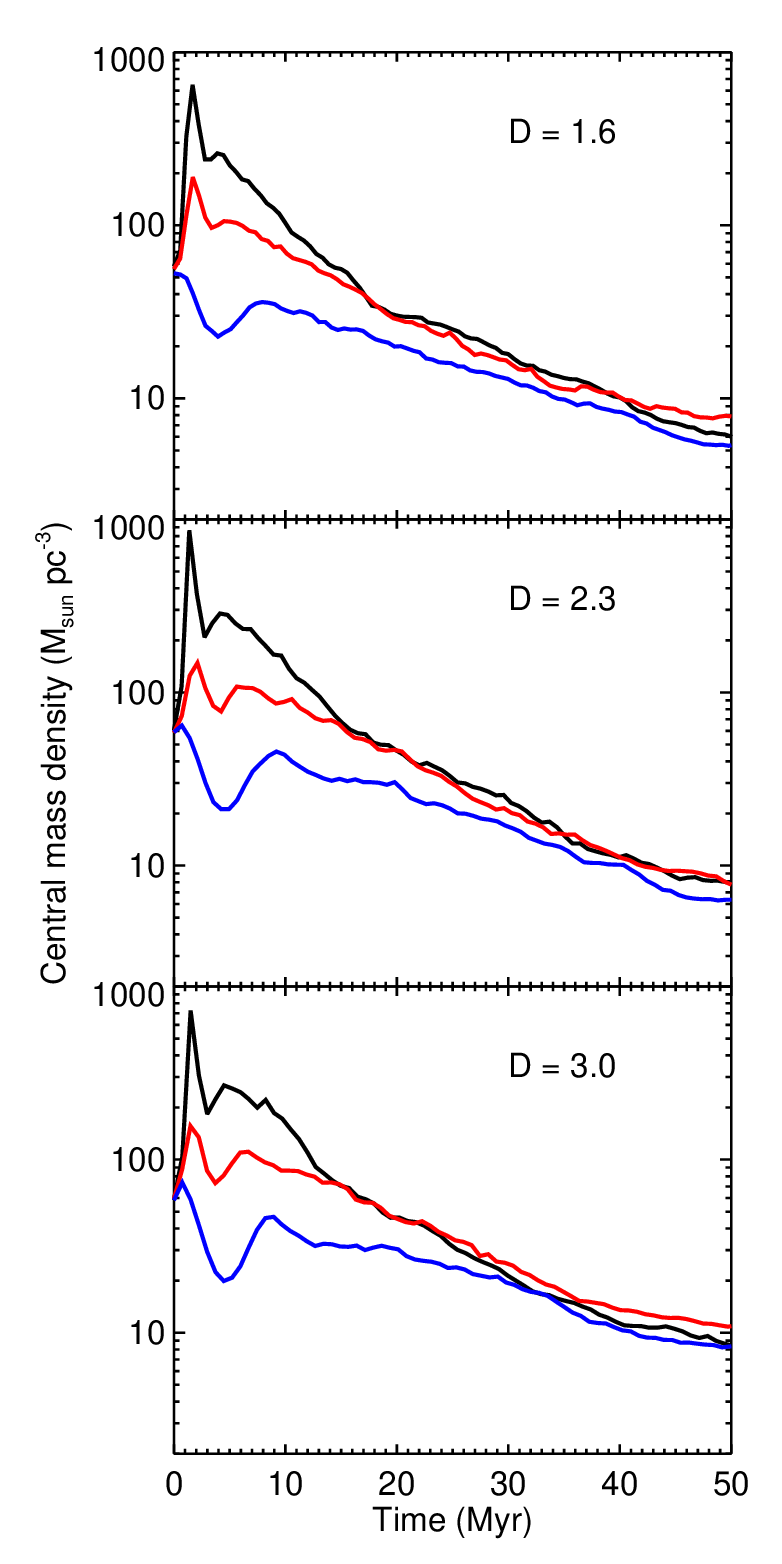} \\
    \end{tabular}
  \caption{The evolution of the half-mass radius $\halfmass$ ({\em left}) and the central mass density $\rhoc$ ({\em right}) for $D = 1.6, 2.3, 3.0$ (averaged over the ensemble of realisations). The different curves in each panel indicate models with initial virial ratios of $Q = 0.3$ ({\em black}), $Q=0.5$ ({\em red}) and $Q=0.7$ ({\em blue}). }
  \label{fig:cluster_infor}
\end{figure}

The dynamical evolution of substructured star clusters has been studied extensively \citep[see, e.g.,][and references therein]{cartwright2004, goodwin2004, allison2009, parkerquanz}. As in previous studies, we find that the initial spatial distribution of stars in each cluster (see, e.g., Fig.~\ref{fig:cluster0}) changes drastically on a short time scale. These changes typically occur within several dynamical timescales, $\tdyn$, which is usually defined as
\begin{equation} \label{eq:dynamicaltime}
  \tdyn \approx 2 \times 10^4~\left(\frac{\mcluster}{10^6 {\rm M}_\odot}\right)^{-1/2} \left(\frac{\halfmass}{{\rm 1~pc}} \right)^{3/2} ~{\rm yr}
\end{equation} 
\citep[e.g.,][]{spitzer1987}. Although our models do not satisfy all necessary conditions for applying this equation, particularly in the highly-substructured clusters, it can be used to estimate a dynamical time scale of $\tdyn\approx 0.8$~Myr in our reference model. 
After substructure has been removed and virial equilibrium achieved, the clusters slowly evolve due to two-body relaxation, stellar evolution, and the effect of the Galactic tidal field. Two-body encounters typically happen within a relaxation time,  
\begin{equation}  \label{eq:relaxationtime}
  \trelax = \frac{0.206 \nstar \halfmass^{3/2}}{\sqrt{G\mcluster}\ln \Lambda} \ ,
\end{equation}
where $G$ is the gravitational constant, $\nstar$ is the number of stars, and $\ln\Lambda \approx \ln \nstar$ is the Coulomb logarithm \citep{binneytremaine, heggiehut}. Although the modelled star clusters are initially substructured, Eq.~\ref{eq:relaxationtime} provides a reasonable first-order estimate for the initial relaxation time, which is roughly $\trelax \approx 18$~Myr. It must be noted, however, that substructured star clusters tend to evolve faster (see Section~\ref{section:method}) and show earlier signs of mass segregation  \citep{allison2009, allison2010} than initially smooth star clusters.

In Fig.~\ref{fig:cluster_infor} we show the evolution of the half-mass
radius $\halfmass$ and the central mass density $\rhoc$ of each star
cluster. Following \cite{parkerquanz} we define the central density as
the average stellar density within the half-mass radius $\halfmass$, i.e.,
\begin{equation} \label{eq:density}
  \rhoc =  \frac{3\mcluster}{8\pi \halfmass^{3}} \ .
\end{equation}
Although this definition of the density allows us to describe the overall evolution of the star clusters, it is still an average, and therefore tends to overestimate or underestimate of the local stellar densities, particularly in highly-substructured star clusters \citep[see, e.g.,][]{parkerdale2013, parkerwright2014, parkerdensity2014}. This results in a higher destruction rate as compared to estimates from Eq.~\ref{eq:density}, particularly in the central region; we will address this issue in Section~\ref{section:dependenced}.

At $t=0$~Myr all star clusters have $\halfmass=1$~pc, as specified by the initial conditions \citep[note that our clusters are typically a factor 1.26 larger than those of][]{parkerquanz}.
The time at which the star clusters experience a gradual transition from substructured systems into a smooth, slowly-evolving systems, is roughly proportional to $Q$, and depends only mildly on $D$.
Apart from the cases with $Q = 0.7$, the clusters initially contract, after which $\halfmass$ increases with time.  
All clusters with $Q=0.3$ reach a central density higher than $700~\msun$\,pc$^{-3}$, which corresponds to an average stellar separation of $10^4$~AU.
The clusters $Q = 0.5$ only reach a modestly high central density of $\rhoc \approx 200          $~stars\,pc$^{-3}$. 
The clusters with $Q = 0.7$ initially expand, until a large number of stars reaches reaches the tidal radius and escape. Due to the large number of escapers for the $Q=0.7$ models, cluster membership is decreased, and the bound stars remain in a cluster with a smaller half-mass radius and larger central density.
For times $t \ga \trelax$, all clusters expand and decrease in central density. Those that are initially supervirial obtain a larger $\halfmass$ and a lower $\rhoc$ than star clusters starting out with $Q=0.5$, while the star clusters with $Q=0.3$ generally have intermediate values for $\halfmass$ and $\rhoc$.

The close encounter probability is roughly proportional to $\rhoc$, which changes by almost two orders of magnitude during the first 50~Myr, although local density variations due to initial substructure may differ substantially. The vast majority of close encounters and planetary disruptions therefore occur during the high-density phase during the first $\sim 20$~Myr. Although initial substructure and the
initial virial state mostly affect the evolution of the
cluster at early times, they play an important role in determining the distribution of the planets over the four dynamical states at later times.


\subsection{Evolution of the planetary population} \label{section:dependencetime}


\subsubsection{Properties of the planet population}

\begin{figure}
 \includegraphics[width=0.45\textwidth]{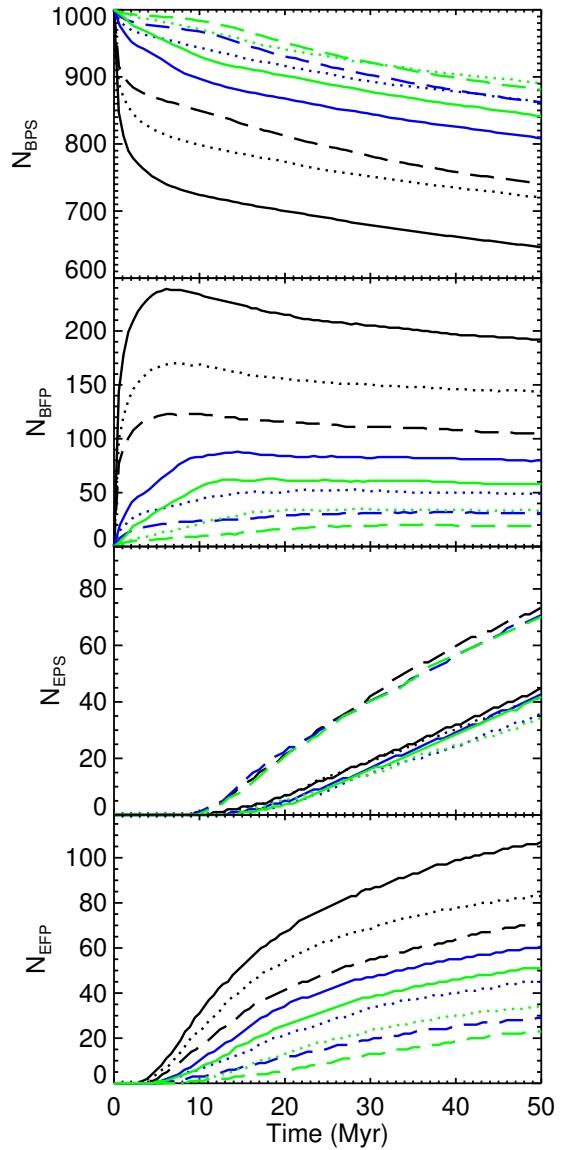}
 \caption{The evolution of $\nbps$, $\nbfp$, $\neps$ and $\nefp$ in the star clusters (averaged of over the ensemble of realisations). All planets initially have $a=100$~AU. Results are shown for star clusters with initial substructure parameters $D=1.6$ (black), $D=2.3$ (blue) and $D=3.0$ (green) and initial virial ratios $Q = 0.3$ (solid curves), $Q=0.5$ (dotted curves) and $Q=0.7$ (dashed curves).  }
   \label{fig:evolution_one}
\end{figure}

 \begin{figure*}
 \includegraphics[width=1\textwidth]{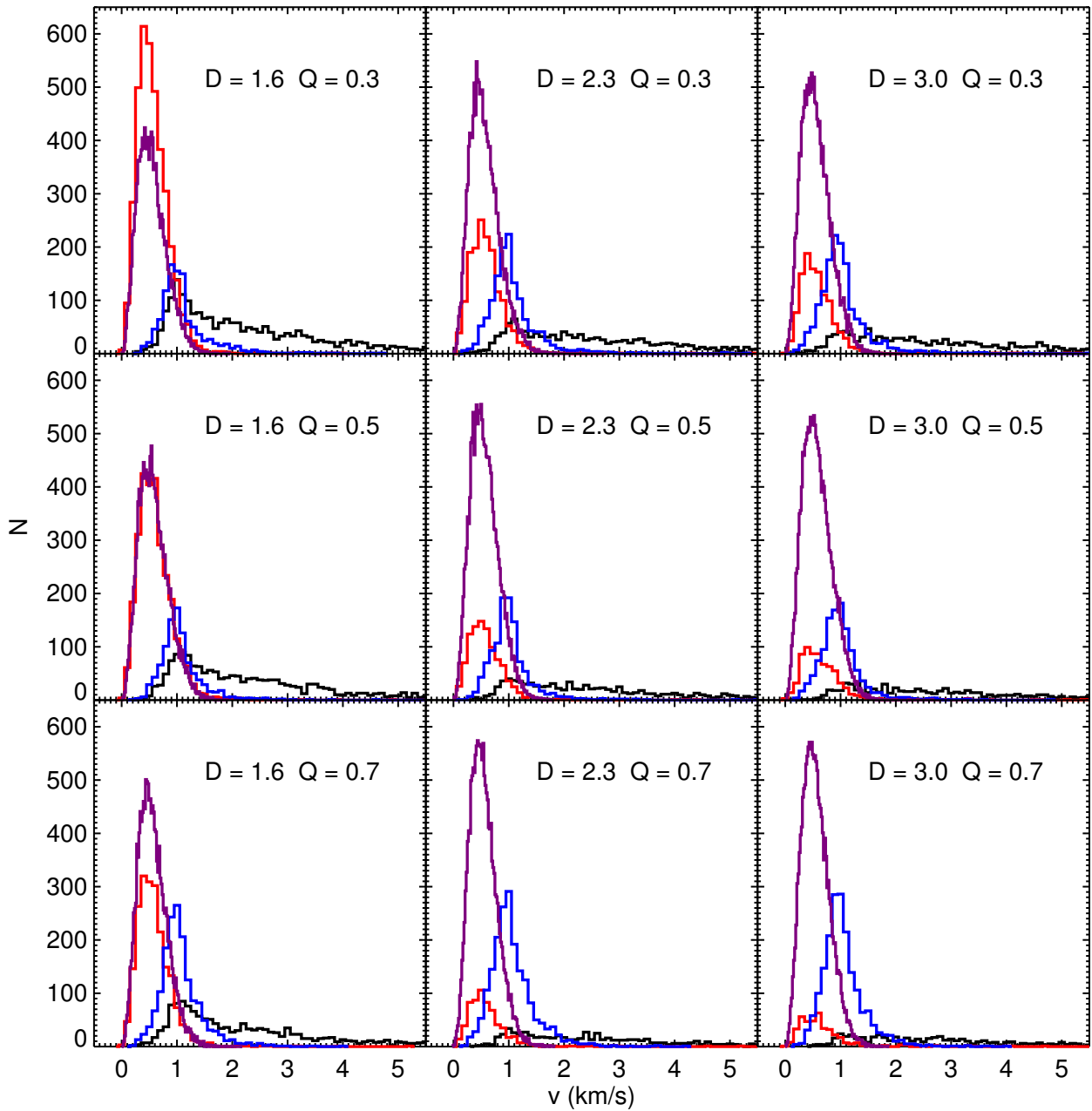}
 \caption{The velocity distribution of BPS (purple), BFP (red), EPS (blue) and EFP (black),
   at $t = 50$~Myr, for the combined sets of twenty realisations.  }
 \label{fig:velocity}
 \end{figure*}

The evolution of $\nbps$, $\nbfp$, $\neps$ and $\nefp$ is shown in Fig.~\ref{fig:evolution_one}, for clusters with different $D$ and $Q$. The planetary population experiences most evolution during the first $\sim 5$~Myr. The most pronounced changes are seen for highly-substructured clusters and those out of virial equilibrium, resulting in smaller $\nbps$ and larger $\neps$, $\nbfp$ and $\nefp$ at any time. During the first $\sim 5$~Myr, $\nbps$ decreases and $\nbfp$ increases, indicating that many of the ionised planets initially remain part of the cluster. A substantial fraction of these BFPs have speeds up to a few \kms{} above the local escape velocity (see below). As planets are marked as escapers beyond two tidal radii, it takes $\delta t \approx 2R_t/v\approx 2-6$~Myr to escape. At later times, after the star clusters have obtained a state of quasi-equilibrium, $\nbps$ and $\nbfp$ both slowly decrease with time due to gradual disruption of BPSs and gradual escape of BPSs and BPFs. The nearly flat distribution $\nbfp(t)$ indicates that the production and  escapes rate of BFPs are roughly equal for $t\ga 10$~Myr.
As dynamical capture of free-floating planets is rare \citep[e.g.,][]{parkerquanz, perets2012}, the increase in $\neps$ is almost entirely due to escape of existing planetary systems (Eq.~\ref{eq:tracks}). As virialised star clusters tend to lose stars at a roughly constant rate \citep[e.g.,][]{heggiehut}, and since none of the planetary systems is disrupted after escape, $\neps$ increases more or less linearly with time. 

As expected, the dynamical evolution is strongest for the models with ($D=1.6$, $Q=0.3$) and weakest for those with ($D=3.0$, $Q=0.7$). At $t=50$~Myr, the ($D=1.6$, $Q=0.3$) clusters have lost 35\% of their BPSs: 4\% have escaped the cluster intact as EPS, 12\% escaped from the cluster as EFPs, and 19\% remain part of the cluster as BFPs. On the other hand, at $t=50$~Myr, the ($D=3.0$, $Q=0.7$) clusters only lose 11\% of their BPS: 7\% become EPSs, 2\% remain in the cluster as BFPs, and 2\% escape as EFPs. 
Clusters with other initial conditions show similar or intermediate behaviour. Variations in $D$ and $Q$ result in a scaling down of $\nbps$, and in a scaling down of the equilibrium value (beyond $t \ga 10$~Myr) for $\nbfp$, and also results in an upward scaling of $\neps$ and $\nefp$. An EFP is generated if and only if a BFP escapes. As free-floating planets tend to escape from star clusters substantially earlier than stellar members \citep[e.g.,][]{parkerquanz, wang2015a}, $\nefp$ increases fast at early times, after which it grows at a steady rate when the BFP equilibrium is established.
As all clusters are smooth and virialised beyond $t\approx 10$~Myr, the differences at later times can be explained as being the result of the initial dynamical evolution of the star clusters under different values of $D$ and $Q$, more specifically, as a result of the variations in stellar density during the early phase of evolution \citep[see, e.g.,][]{parker2011}.

Fig.~\ref{fig:velocity} shows the velocity distributions at 50~Myr. Those of the BPSs and BFPs are more or less identical, although BFPs tend to roam around in the outskirts of the cluster and these therefore have slightly lower velocities. 
The escapers (EPS and EFP) have higher velocities, usually $1-3$~\kms{} above the escape velocity, consistent with the findings of \cite{parkerquanz}. Some EFPs reach velocities as high as $\sim 30$~\kms{} and escape as "runaway planets" following a strong dynamical interaction.  
There is no strong correlation between the velocity distribution of the EPSs and the initial conditions.  
 

\begin{table*}
   \centering
   \caption{The semi-major axis $a$ and eccentricity $e$ of the remaining planetary systems at $t=50$~Myr, for star clusters with different initial substructure parameters $D$ and virial ratios $Q$. The "$\approx$" symbol indicates changes of less than one part per million. The percentages represent the averages of twenty realisations. The range of values for the ensemble of simulations is indicated between the brackets. }
     \begin{tabular}{ccccccccccccc}
       \hline \hline
       $D$ & $Q$   &  \multicolumn{2}{c}{$a\approx100$~AU, $e\approx0$}  &   \multicolumn{2}{c}{$a\approx100$~AU,  $e>0$}   &   \multicolumn{2}{c}{$a>100$~AU,  all $e$}  &  \multicolumn{2}{c}{$a<100$~AU,  all $e$}  & \multicolumn{2}{c}{all $a$, $e\ge 0.1$}  \\
           &       &    \multicolumn{2}{c}{per cent}       &    \multicolumn{2}{c}{per cent}          &  \multicolumn{2}{c}{per cent}          & \multicolumn{2}{c}{per cent}       &  \multicolumn{2}{c}{per cent}   \\
       \hline
1.6   &   0.3   &   4.4 & $(0.0-28.6)$    &   4.9 & $(1.3-21.7)$    &   17.7 & $(12.0-22.7)$    &   72.6 & $(30.1-84.7)$    &   24.7 & $(17.6-32.4)$  \\
2.3   &   0.3   &   0.6 & $(0.0-1.5)$    &   2.5 & $(0.8-5.7)$    &   15.6 & $(13.4-20.3)$    &   81.2 & $(77.9-85.8)$    &   19.4 & $(16.4-24.4)$  \\
3.0   &   0.3   &   0.7 & $(0.0-1.7)$    &   2.2 & $(0.9-4.0)$    &   14.6 & $(11.0-18.1)$    &   82.5 & $(77.5-87.5)$    &   17.6 & $(12.6-22.5)$  \\
\hline
1.6   &   0.5   &   0.7 & $(0.0-1.6)$    &   1.8 & $(0.6-5.0)$    &   12.3 & $(9.3-17.8)$    &   85.1 & $(80.4-88.8)$    &   17.3 & $(12.0-22.6)$  \\
2.3   &   0.5   &   0.9 & $(0.2-1.6)$    &   1.2 & $(0.2-3.3)$    &   10.3 & $(7.3-13.3)$    &   87.6 & $(84.6-90.6)$    &   12.9 & $(9.4-16.8)$  \\
3.0   &   0.5   &   0.7 & $(0.2-1.4)$    &   1.2 & $(0.0-2.0)$    &   10.4 & $(8.5-16.4)$    &   87.7 & $(81.4-90.7)$    &   11.4 & $(7.3-14.6)$  \\
\hline
1.6   &   0.7   &   1.1 & $(0.0-7.2)$    &   1.5 & $(0.3-4.9)$    &   10.2 & $(6.0-14.2)$    &   87.2 & $(73.6-93.7)$    &   13.6 & $(9.4-19.1)$  \\
2.3   &   0.7   &   0.8 & $(0.2-1.6)$    &   0.8 & $(0.0-2.0)$    &   9.0 & $(6.7-12.4)$    &   89.4 & $(84.2-91.8)$    &   9.1 & $(4.7-11.9)$  \\
3.0   &   0.7   &   1.1 & $(0.2-2.4)$    &   1.2 & $(0.2-2.0)$    &   8.6 & $(5.9-11.2)$    &   89.1 & $(85.5-91.8)$    &   7.4 & $(5.8-9.5)$  \\
       \hline \hline
       \label{tab:2}
     \end{tabular}
 \end{table*}
 
Non-disruptive dynamical encounters tend to alter the orbital elements of BFSs, as shown in \cite{spurzem2009} and \cite{parkerquanz}. In Table~\ref{tab:2} we summarise the semi-major axis $a$ and eccentricity $e$ distributions for our clusters at $t=50$~Myr. Weak encounters tend to result in more angular momentum exchange (changes in $e$ and $i$) than energy exchange (changes in $a$), resulting in the familiar fountain-diagrams \citep[e.g., figure~5 in][]{parkerquanz}. Eccentricity growth is largest for highly substructured, subvirial clusters, where dynamical interactions are frequent. Semi-major axes remain mostly unchanged, although the fraction of softened BPSs (larger $a$) is again largest for clusters with $D=1.6$ and $Q=0.3$, as close encounters in these clusters are most frequent. On average, $70-90\%$ of the BPSs that remain at $t=50$~Myr have slightly hardened (smaller $a$), and this fraction is largest for the least violent clusters. It should be noted, however, that most of the hardened orbits experience a minimal decrease in semi-major axis, while the softened orbits can become substantially wider \citep[see, e.g., figure~5 in][and note that the horizontal axis is logarithmic]{parkerquanz}.
 

 \subsubsection{The fraction of planet-hosting stars in star clusters and in the field} \label{section:planethosting}

   \begin{figure*}
 \includegraphics[width=0.8\textwidth]{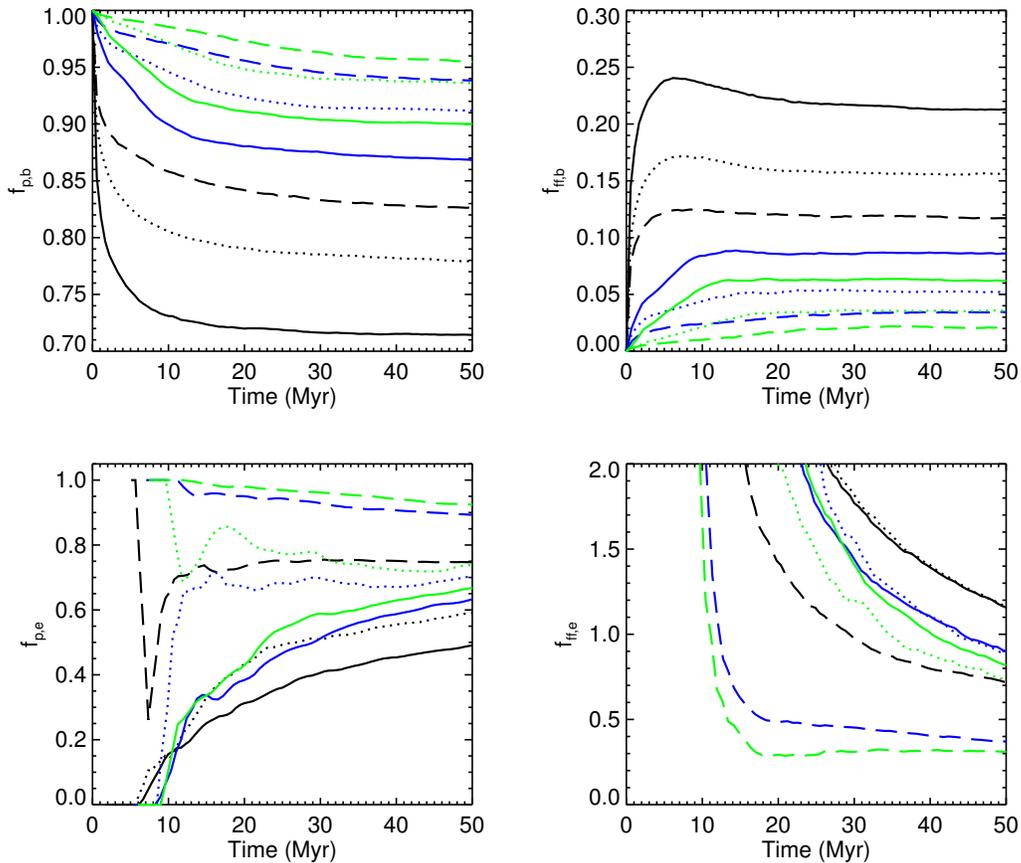}
 \caption{The fraction of stars with planets $\fplanetbound$ in the star cluster ({\em top-left}), the fraction of stars with planets $\fplanetunbound$ among the escaping stars ({\em bottom-left}), the ratio between free-floating planets and stars in the cluster $\fffbound$ ({\em top-right}), and the ratio between free-floating planets and stars among the escaped objects $\fffunbound$ ({\em bottom-right}) as a function of time (Eqs.~\ref{eq:fplanetbound} and~\ref{eq:ffbound}). Results are shown for star clusters with initial substructures $D = 1.6$ (black), $D = 2.3$ (blue) and $D = 3.0$ (green) and initial viral parameters $Q = 0.3$ (solid curves), $Q = 0.5$ (dotted curves) and $Q = 0.7$ (dashed curves).   }
   \label{fig:evolution}
 \end{figure*} 

In the previous sections we described the dynamical properties of the entire set of planets. Dynamically, this makes sense, as we can follow all planets in our simulations. Observationally, it is more useful to consider the fraction of planet-hosting stars in the cluster and the field, and the number of free-floating planets in the cluster and in the field, with respect to the number of stars in the same environment.
At a given time, the fraction of planet-hosting stars $\fplanetbound$ in the star cluster and $\fplanetunbound$ in the field are
\begin{equation} \label{eq:fplanetbound}
  \fplanetbound = \frac{\nbps}{\nbs+\nbps} \quad {\rm and} \quad \fplanetunbound = \frac{\neps}{\nes+\neps} \ ,
\end{equation}
respectively. Here, $\nbs$ and $\nes$ represent the number of stars without a planetary companion in the star cluster and in the field, respectively. Similarly, the ratio between the number of free-floating planets and stars among the bound and unbound objects is
\begin{equation} \label{eq:ffbound}
  \fffbound = \frac{\nbfp}{\nbs+\nbps} \quad {\rm and} \quad \fffunbound = \frac{\nefp}{\nes+\neps} \ .
\end{equation}
Since all planets initially orbit a cluster member, the initial values are $\fplanetbound=1$, $\fffbound=0$, while $\fplanetunbound$ and $\fffunbound$ are undefined. 

Fig.~\ref{fig:evolution} shows $\fplanetbound$, $\fplanetunbound$, $\fffbound$, and $\fffunbound$ as a function of time for clusters with different initial $Q$ and $D$. A comparison between Figures~\ref{fig:evolution_one} and~\ref{fig:evolution} shows a strong correlation between $\fplanetbound$ and $\nbps$, which is not surprising since the number of stars without a planet in the star cluster ($\nbs$ in Eq.~\ref{eq:fplanetbound}) is generally much smaller than $\nbps$. During the initial virialisation process the fraction $\fplanetbound$ rapidly decreases, after which it obtains its equilibrium at a value that depends on $D$ and $Q$. The ratio $\fplanetunbound$ for the escaped stars is undefined during the first $\sim 5$~Myr, as no star has had the time to escape from the cluster. Subsequently, $\fplanetunbound$ grows with time (after a short period strong fluctuations due to low-number statistics), because BFPs tend to escape at earlier times than EPS (see Fig.~\ref{fig:evolution_one}) as on average they obtain higher velocities after experiencing close encounters with other cluster members. The fractions $\fplanetbound$ and $\fplanetbound$ show a similar dependence on $D$ and $Q$.

As $\nbs$ is small compared to $\nbps$ at any time, the evolution of $\fffbound$ also correlates strongly with $\fbps$ (Fig.~\ref{fig:evolution_one}). The ratio $\fffunbound$ among the escaping objects is initially undefined, and as initially only free-floating planets escape from the cluster, the value is initially above unity. At later times, stars also escape and the values drop below unity. For the models with $Q=0.7$ the ratio $\fffunbound$ drops to low values (more stars than EFPs), as most planetary systems escape intact (as EPSs), while for highly substructured and subvirial star clusters, $\fffunbound$ remains above unity (more EFPs than stars) at $t=50$~Myr.


 \subsection{Dependence on initial semi-major axis $a$} \label{section:dependencesma}

\begin{figure*}
 \includegraphics[width=1\textwidth]{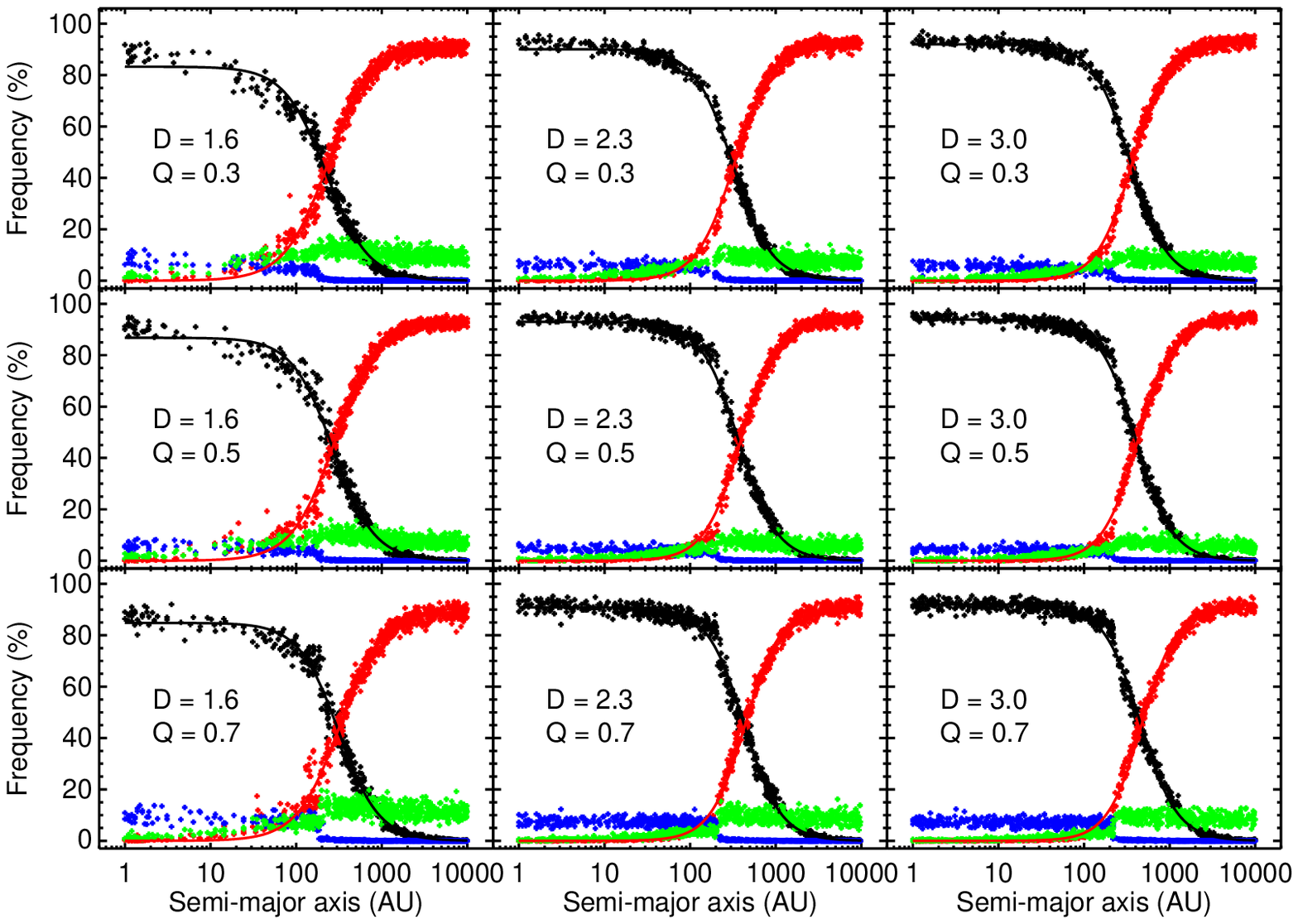}
 \caption{The frequency of BPS (black), BFP (red), EPS
   (blue), and EFP (green) at $t= 50$~Myr,
   as a function of initial semi-major axis. The black and red curves are well fitted by the distribution $f(a)=f_0 (1+[a/a_0]^c)^{-1}$, and the fitted values are listed in Table~\ref{table:fittedvalues}.}
 \label{fig:6}
 \end{figure*}

 \begin{table}
   \caption{The fitted values for the distribution $f_0 (1+[a/a_0]^c)^{-1}$ for bound planetary systems and bound free-floating planets in Fig.~\ref{fig:6}, at $t=50$~Myr.  \label{table:fittedvalues} }
   \begin{tabular}{cc ccc ccc}
     \hline \hline
     $D$ & $Q$ &       & BPS   & & &  BFP  &     \\ 
     && $f_0$ & $a_0$ & $c$ & $f_0$ & $a_0$ & $c$ \\
     \hline
1.6 & 0.3 & 83.3 & 226.0 & 1.77 & 91.3 & 235.9 & --1.69 \\ 
2.3 & 0.3 & 90.1 & 317.2 & 1.94 & 92.7 & 334.5 & --1.95 \\ 
3.0 & 0.3 & 92.0 & 348.5 & 2.02 & 92.9 & 370.0 & --2.07 \\ 
\hline
1.6 & 0.5 & 86.8 & 270.7 & 1.81 & 93.3 & 283.5 & --1.73 \\ 
2.3 & 0.5 & 93.0 & 371.5 & 1.97 & 94.2 & 393.9 & --2.03 \\ 
3.0 & 0.5 & 93.8 & 405.7 & 2.02 & 94.3 & 427.2 & --2.10 \\ 
\hline
1.6 & 0.7 & 84.8 & 315.3 & 1.85 & 89.3 & 322.9 & --1.74 \\ 
2.3 & 0.7 & 90.8 & 426.1 & 2.03 & 90.6 & 440.4 & --2.13 \\ 
3.0 & 0.7 & 92.0 & 454.0 & 2.02 & 90.5 & 470.5 & --2.17 \\ 
     \hline \hline
   \end{tabular} 
\end{table}

\begin{figure}
 \includegraphics[width=0.5\textwidth]{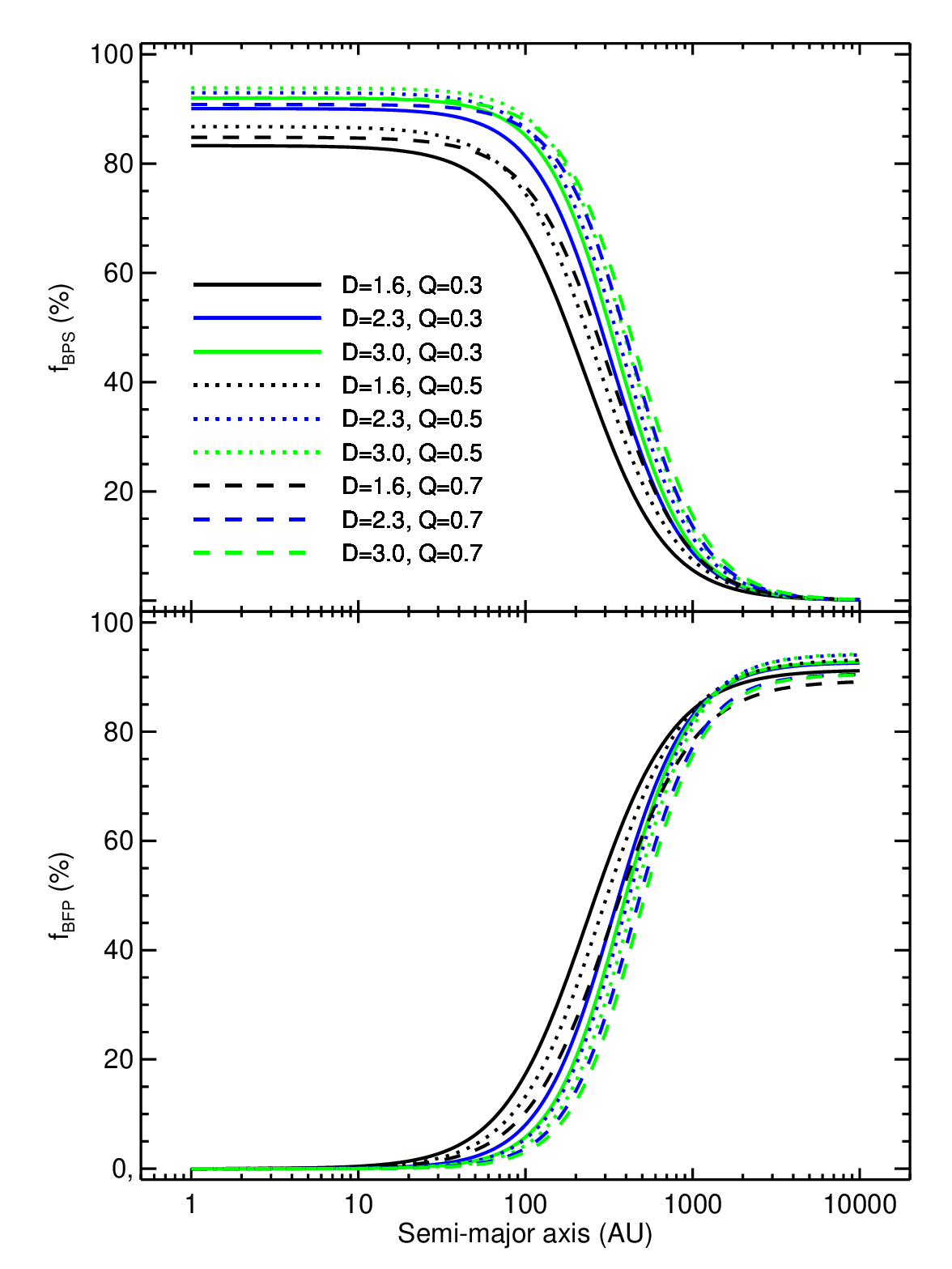}\\
 \caption{Fitted curves describing the frequency $\fbps$ ({\em top}) and $\fbfp$ ({\em bottom}) at $t=50$~Myr, as a function of semi-major axis (cf. Fig.~\ref{fig:6}).}
 \label{fig:fit_p}
 \end{figure}

The survival chances of planetary systems depend on both their environment and on their initial binding energy, $E_{b} \propto a^{-1}$. We study the latter dependence by modelling the evolution planetary systems over a large range of semi-major axes, 1~AU\,$ \le a \le 10\,000$~AU, and study their evolution for an ensemble of eight hundred star clusters.
Fig.~\ref{fig:6} shows the dynamical fate of planets with different semi-major axes at $t=50$~Myr for the different clusters. 

Most planetary systems with $a \la 100$~AU survive as star cluster members, although a small fraction escape as EPSs. Almost all systems with $a \ga 2000$~AU are destroyed and the majority of these remain as BFPs in the star cluster, while some escape as EFPs. Survival chances drop sharply from $a \approx$ 100~AU to $a \approx 2000$~AU, and as a result, $\fbfp$ increases and $\fbps$ decreases rapidly with increasing $a$. As discussed earlier (see Fig.~\ref{fig:evolution_one}), most of these changes occur during the first $\sim 5$~Myr. The fraction of planetary systems that escape the cluster intact (EPS; blue symbols) is typically $5-15\%$ when $a \la 200$~AU. The number of EFPs slowly increases with increasing $a$, as more systems are disrupted. The number of EFPs also drastically increase around $a\approx 200$~AU. 

In order to quantify our results, we fit $\fbps(a)$ and $\fbfp(a)$ to distinguish between the effects of different initial conditions in Fig.~\ref{fig:6}. All curves are well described with a function
 \begin{equation} \label{eq:goodfit}
   f(a) = f_0 \left[ 1+ \left(\frac{a}{a_0}\right)^c\right]^{-1} \ ,
 \end{equation}
 where $f_0, a_0$ and $c$ are constants. For $a \ll a_0$, we obtain $\fbps(a)+\feps(a)
 \approx 1$ and $\fbfp(a)=\feps(a)\approx 0$, which refers to the initial conditions. 
 For $a \gg a_0$
 we obtain $\fbps(a)+\feps(a) \approx 0$, indicating that almost all systems are
 destroyed. 
The fitted quantities $f_0$, $a_0$ and $c$ of the BPS and BFP populations are listed in Table~\ref{table:fittedvalues} for models with different initial $D$ and $Q$. For comparison, the fits for $\fbps(a)$ and $\fbfp(a)$ are also combined in Fig.~\ref{fig:fit_p}. 

Strong transitions occur in the region around $a= a_0$, which is the semi-major axis boundary that separates the domains in which planetary systems remain mostly intact or are mostly destroyed. The quantity $|c|$ indicates how broad this region transition is in terms of semi-major axis. Note that a smaller value of $|c|$ corresponds to a broader transition region.  Note the value of $a_0$ is unrelated to the hard-soft boundary that is often used to describe the dynamics of binary systems in star clusters; all planets are weakly bound to their host star and are therefore by definition "soft". The destruction of planetary systems is determined by the close encounter rate, rather than by the encounter energy.

For the BPSs, the quantity $f_0 \approx 83-94\%$ indicates the fraction of stars remaining in the star cluster. Since almost all planetary systems remain intact when $a \ll a_0$, $f(a)\approx f_0$ for the BPS and $f(a)=1-f_0$ for the EPS in this regime. In other words, $6-17\%$ of the planetary systems escape the cluster within 50~Myr. As described earlier the number of planetary systems that remain part of the star cluster ($f_0$)  increases with increasing degree of homogeneity, and is largest for clusters starting out in virial equilibrium.
In the limit $a\gg a_0$, virtually all planetary systems are destroyed, and free-floating planets either remain in the cluster as BFPs or escape as EFPs. In this limit, $f(a)=f_0 \approx 91-94\%$ for the BFPs, and $f(a)=1-f_0=6-9\%$ for the EPSs. This demonstrates that even when planetary systems are destroyed at  early times, most free-floating planets remain a member of their host star cluster during the first $\sim 50$~Myr. 
The critical semi-major axis $a_0$ is in the range $200-500$~AU, with the exact value depending on the initial conditions. BPSs in more substructured star clusters experience more close encounters, and therefore have a smaller $a_0$. Planetary systems have the largest chances of being retained (i.e., no disruption or escape) in the least violent clusters ($Q=0.7$ and $D=3.0$), and these clusters thus have the largest $a_0$. 

The width in the transition region is described by the parameter $|c|$, where the sign of $c$ determines whether the distribution increases or decreases with $a$. Part of this variation is a result of the global density variations in the star clusters, and part can be attributed to the variation in binding energies (host stars with different stellar masses). For the fits of both the BPS and the BFP, $|c|$ increases with increasing $D$, indicating that the transition range (around $a_0$) for destruction of planetary systems is broader. This is understandable, since an increased amount of initial substructure results in a larger variation in local stellar densities and therefore also encounter frequencies. For the BPSs, $|c|$ depends only mildly on $Q$, while for the BFPs, $|c|$ increases with increasing $D$, for the same reason as above.

\cite{bonnell2001} obtained comparable results using analytical estimates (see their figures~4 and~5). They study the evolution of planetary systems in different environments over much longer time scales (up to 10~Gyr), and focus on planetary semi-major axes in the range $0.1-100$~AU. For the cluster densities considered in our work, only the widest planetary orbits in this separation range are affected at $t=50$~Myr (Fig.~\ref{fig:6}), which is indeed what \cite{bonnell2001} have found. Further $N$-body simulations spanning a much larger range in cluster density and a much longer integration time are required to validate the other analytical results presented in \cite{bonnell2001}.


 \subsection{Dependence on the substructure parameter $D$} \label{section:dependenced}
 
 \begin{figure}
     \includegraphics[width=0.5\textwidth]{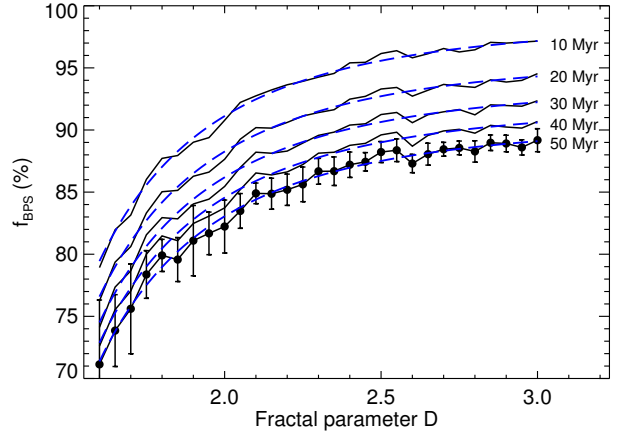} 
   \caption{The dependence of $\fbps(D)$ on the initial substructure parameter $D$. Shaded regions bracket, from dark to light, $\fbps(D)$ at $t=10$, 20, 30, 40, and 50~Myr. The dots represent the median values (and corresponding standard deviations) at $t=50$~Myr, and the solid curve represents a fit to these data (Eq.~\ref{eq:fd}).  }
  \label{fig:f_d}
 \end{figure}

The encounter rate $\Gamma(D)$ of stars in a star cluster with substructure parameter $D$ can, to first order, be approximated with 
$ \Gamma(D) = n(D) \pi p^2_{\rm max} v(D)$, where, $p_{\rm max}$ is a maximum impact parameter that is considered as encounter, $n$ is the local
stellar density and $v(D)$ is the relative velocity at infinity
between two bodies. 
For highly substructured star clusters, $n(D)$ and $v(D)$ vary strongly with location due the presence of high-density pockets of stars, a phenomenon that was discussed earlier by, e.g., \cite{parkerdale2013, parkerwright2014} and \cite{parkerdensity2014}. In those cases, a larger number of planetary systems is disrupted, resulting in a larger $\fbps$ and a smaller $\fbfp$.

To quantify the correlation between $D$ and $\fbps$, we run simulations with thirty different initial substructure parameters in the range $D\in [1.6-3.0]$. All cluster are initially in virial equilibrium ($Q=0.5$), and the remaining initial conditions are identical to those of our reference model (Table~\ref{table:initialconditions}). In order to reduce statistical fluctuations, we run twenty realisations for each value of $D$.
The resulting values $\fbps(D,t)$ are shown in Fig.~\ref{fig:f_d} for $t=10$, 20, 30, 40, and 50~Myr. As can also be seen in Fig.~\ref{fig:evolution_one}, $\fbps$ decreases roughly linearly with time, after the initial phase of relaxation when substructure is removed and virial equilibrium is restored. For the clusters in our sample, the fraction $\fbps(D,t)$ is well described by
 \begin{equation} \label{eq:substructure}
    \fbfp(D,t) \approx A(t) + B(D) \ ,
   \label{eq:fd}
 \end{equation}
where the (nearly linear) evolution over time can be expressed as
 \begin{equation} \label{eq:substructurea}
A(t) \approx 1 -(2.06\times 10^{-3} )t + (3.28\times 10^{-6})t^2 \ ,
 \end{equation}
 where $t$ is in units of Myr. The dependence on the initial substructure parameter as
 \begin{equation} \label{eq:substructureb}
 B(D) \approx \left( 0.78-D^{3.8} \right)^{-1}
 \end{equation}
The approximation in Eq.~\ref{eq:substructure} has of one term that depends only on $t$ and another that only depends on $D$. This means that changes over time, $d\fbps/dt=dA(t)/dt$, are mostly independent of the initial amount of substructure, which can be seen in Fig.~\ref{fig:f_d}. It also means that a different amount of initial substructure does not affect its evolution over time, $d\fbps/dD=dB(D)/dD$, but merely establishes its normalisation at later times. At any moment in time beyond several initial relaxation times, $\fbps$ depends weakly on the initial amount of substructure for $D \ga 2$, when $d\fbps/dD\approx 0$, which explains the similarities in Fig.~\ref{fig:fit_p}. Note that, as expected, $\fbps(D,t=0)\approx 1$ corresponds to our initial conditions for $D\ga 2$. For $D<2$, $\fbps(D,t=0)> 1$, which shows that our simple approximation is not valid during the earliest phases of evolution of star clusters with a high amount of initial substructure.


 \subsection{Dependence on initial mass and density}\label{section:dependenceclustermass}

\begin{figure}
 \includegraphics[width=0.5\textwidth]{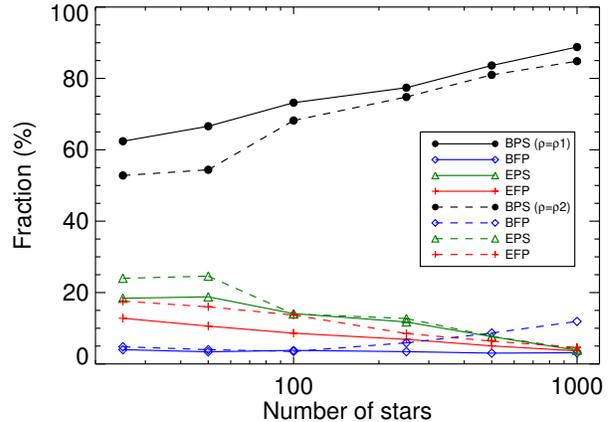}
 \caption{The frequencies $\fbps$, $\feps$, $\fbfp$ and $\fefp$ at $t=50$~Myr, as a function of the number of stars in the cluster $\nstar$. Results are shown for two clusters with central stellar densities (Eq.~\ref{eq:density}) $\rho_1=120$~pc$^{-3}$ and $\rho_2=280$~pc$^{-3}$.}
   \label{fig:f_N}
 \end{figure} 

In the previous sections we discussed the evolution of star clusters that all consisted of an identical number of stars ($\nstar=1000$) and had an identical size ($\halfmass=1$~pc). In order to broaden the applicability of our results, we discuss in this section how the evolution of the planetary population depends on the initial mass and central density of the star clusters. 

We carry out simulations with $\nstar = 25$, 50, 100, 250, 500, and 1000 stars. In order to allow a fair comparison between the clusters of different mass, one would ideally keep all other star cluster parameters (such as its half-mass radius $\halfmass$, its density $\rho$, and its half-mass relaxation time $\trelax$) constant. As all these parameters depend on $\nstar$ and on each other, only combinations of parameters can be kept constant \citep[see][for an extensive discussion on how different combinations of these parameters affect star cluster evolution]{kouwenhoven2014}. As we are mostly interested in dynamics, we compare star clusters in which the initial stellar density is constant for all values of $\nstar$.
We model two sets of star clusters: one set in which all star clusters have central stellar densities (Eq.~\ref{eq:density}) of $\rho_1=120$~pc$^{-3}$ (which includes our reference model) and another set in which all clusters have $\rho_2=280$~pc$^{-3}$. Consequently, the corresponding half-mass radii can be expressed as $\halfmass$$_1$$ = 0.1 \nstar^{1/3} $~pc and $\halfmass$$_2$$= 0.075 \nstar^{1/3} $~pc. All star clusters initially have $D=3.0$ and $Q=0.5$, and all stars initially host a planet in a circular orbit at $a=100$~AU. The fraction of planets in the different dynamical categories at $t=50$~Myr are shown in Fig.~\ref{fig:f_N}, and results are obtained by averaging ten realisations for each model.

The (initial) dynamical time scales as $\tdyn \propto \rho^{-1/2}$ (Eq.~\ref{eq:dynamicaltime}), and is therefore identical for star clusters over a large mass range, with a given initial stellar density, corresponding to $\tdyn \approx 0.82$~Myr for $\rho_1$ and $0.53$~Myr for $\rho_2$. 
The relaxation time, on the other hand, scales approximately as $\trelax \propto \nstar\rho^{-1/2}$ (Eq.~\ref{eq:relaxationtime}), and ranges between 1~Myr to 18~Myr for the least and most massive clusters in our sample, respectively. The dynamical age of the star clusters at $t=50$~Myr is thus proportional to $\trelax \propto\tdyn/\nstar$  and is independent of $\rho$. Finally, the tidal radius (Eq.~\ref{eq:tidalradius}) of a cluster scales as $R_t \propto N^{1/3}$, and is also independent of $\rho$. 

To understand dependence on the initial density in Fig.~\ref{fig:f_N}, it is convenient to first consider the evolution of the stellar population alone, and ignore the presence of planets, which have a negligible effect on the stellar dynamics. Clusters with a given $\nstar$ are all evolved for an identical number of relaxation times, despite the different stellar densities. The only relevant difference between the clusters is the tidal radius, more specifically the ratio $\halfmass/R_t\propto\rho^{-1}$. Clusters with a larger initial density are more compact, and it therefore takes longer for BPSs and single stars to escape while the clusters gradually expand to fill their tidal radii. Consequently, clusters with larger $\rho$ have a larger number of bound stars, after an identical number of dynamical times have elapsed. Larger densities, however, do result in the destruction of more planetary systems, and therefore exhibit a larger fraction of BFPs and EFPs at 50~Myr.

Fig.~\ref{fig:f_N} shows that $\fbps$ increases with increasing star
 cluster mass. In general, $\fbps$ on both the
 stellar density, and the binding energy of the star cluster. The
 former is related to the strength and frequency of close encounters,
 so for the two different star density cases, it is reasonable to see
 that when a cluster has a higher initial density, it has smaller
 $\fbps$ at $t=50$~Myr. Due to the low cluster masses, many of the liberated planets are ejected from their host stars with velocities above the cluster's escape velocity, and therefore $\fbfp$ is close to zero and $\fefp$ is positive (apart from the higher-mass clusters that can retain free-floating planets). Planetary systems escaping the cluster intact are most common among the lowest-mass clusters that experience near-dissolution around $t=50$~Myr.


\section{Conclusions and discussion} \label{section:conclusions}

Although many surveys have focused on detecting planets in open and globular clusters, relatively few planets orbiting stars and free-floating planets have been discovered in these environments. This can partially be attributed to observational difficulties, but also to disruption following close encounters with neighbouring stars and subsequent escape of free-floating planets from star clusters. Inspired by the earlier study of \cite{parkerquanz} we carry out $N$-body simulations of evolving star clusters. We employ the \texttt{NBODY6} package to model open star cluster environments with $\nstar=1000$
stars, in which we assign each star a Jupiter-mass companion. All
planets initially have circular and randomly oriented orbits.  
We evolve all clusters for 50~Myr and study how the evolution of the planetary population depends on the initial conditions.  We study star clusters with different initial amounts of substructure ($D=1.6-3.0$), different initial virial ratios ($Q=0.3-0.7$) and initial planetary semi-major axes in the range 1~AU$\leq a \leq 10\,000$~AU. To broaden the applicability of our results, we also carry out simulations of star clusters with different total masses and stellar densities. Our main results can be summarised as follows:
\begin{enumerate}

\item The initial values of $D$ and $Q$ mostly affect the evolution of
   planetary systems during the first $\sim 5$~Myr, which is the time
  at which the initial substructure is removed and virial equilibrium
  is achieved. Subsequently, the stellar and planetary populations evolve
  gradually, more or less independent of the initial conditions. Although the
  initial choices for $D$ and $Q$ take effect at early times, their
  influence on the fate of planetary systems can be
  substantial. The amount of disruption of planetary systems and escape of free-floating planets is larger when a star cluster is initialised with a larger amount of substructure or is further away from virial equilibrium. 
  
\item Although both play a role in the disruption rate of planetary systems and the escape rate of planetary systems from the star cluster, the initial amount of substructure, $D$, mostly affects the former, while the initial virial state, $Q$, has most influence on the latter.

\item In addition to environmental factors, the disruption rate of planetary
  systems is strongly correlated with initial semi-major axis $a$. The
  fractions of bound planetary systems $\fbps(a)$ and bound free-floating planets $\fbfp(a)$ are well-described by
  the functional form $f(a)= f_0(1 + (a/a_0)^c)^{-1}$, where $f_0$, $a_0$ and $c$ are constants. In the case of $\fbps$, $f_0$ represents the fraction of stars that remains bound the cluster, $a_0$ is the stability limit for disruption of planetary systems, and $c$ measures the width of the transition region, which is negative for $\fbps(a)$ and positive for $\fbfp(a)$.  

\item A higher degree of initial substructure in the star clusters results in a higher disruption rate of planetary systems. The fraction of bound planetary systems over time is well approximated with $\fbfp(D,t)=A(t)+ B(D)$, where $A(t)$ is an almost linear function of time (Eq.~\ref{eq:substructurea}), and $B(D)$ is a function that increases with $D$ and flattens off when the level of substructure approaches $D=3$ (Eq.~\ref{eq:substructureb}). 
  
\item For clusters with a fixed initial density, the fraction of planetary systems present at $t=50$~Myr increases with increasing cluster mass,
  while the fraction of escaping free-floating planets decreases. Crowded
  environments result in more frequent encounters, so more free-floating
  planets are generated in the high-density centre of the cluster, and
  fewer intact planetary systems are ejected from the star cluster.  
  
\end{enumerate}

Our study has focused on the general dynamical evolution of star-planet systems and free-floating planets in star clusters. Our models represent a simplification of reality, as we have exclusively modelled single-planet systems. Multi-planet systems are substantially more fragile than single-planet systems. Small perturbation of an outer planet can induce strong gravitational interactions with other planet(s) in the system, which may result in a reconfiguration of the system, in the ejection of one or more planets, or in a physical star-planet or planet-planet collision \citep[see, e.g.,][]{hao2013, shara2014}.
Moreover, we have not included primordial binaries.
Observations have indicated that stars of all masses and ages are often part of a binary or multiple stellar system \citep[see, for example,][and numerous others]{duquennoy1991, kouwenhoven2005, kouwenhoven2007, connelley2008a, connelley2008b, raghavan2010, bergfors2010, janson2014, tokovinin2014a, tokovinin2014b}, and that a considerable fraction of the known exoplanets are part of a multi-planet system \citep[e.g.,][]{latham2011}. 
The presence of binary and multiple stellar systems increases encounter rates due to their larger collisional cross-section, and also extends the longevity of star clusters. Although dynamical binary systems form in our modelled star clusters, primordial binary systems are not included at this stage.

The simplifications mentioned above warrant a further dynamical study in which stellar and planetary multiplicity is taken into account.
It is technically very difficult \citep[but not impossible; see][]{malmberg2011,davies2013,cai2015b}
to model the evolution of multi-planet systems in star clusters due to the enormous dynamical ranges in size, time, and mass. However, with the recent upgrade of NBODY6++ \citep[NBODY6++GPU; see][]{wang2015b} and its integration into the AMUSE framework \citep[][]{portegies2013, pelupessy2013, cai2015a}, direct $N$-body simulations of multi-planet systems in massive star clusters may be carried out in the near future.


\section*{Acknowledgments}

We wish to thank the anonymous referee for her/his constructive comments that helped to improve the manuscript considerably. X.C.Z. and L.W. were supported by the Kavli Institute for Astronomy and Astrophysics and Department of Astronomy at Peking University. M.B.N.K. was supported by the Peter and Patricia Gruber Foundation through the PPGF fellowship, by the Peking University One Hundred Talent Fund (985), and by the National Natural Science Foundation of China (grants 11010237, 11050110414, 11173004). This publication was made possible through the support of a grant from the John Templeton Foundation and National Astronomical Observatories of Chinese Academy of Sciences. The opinions expressed in this publication are those of the author(s) do not necessarily reflect the views of the John Templeton Foundation or National Astronomical Observatories of Chinese Academy of Sciences. The funds from John Templeton Foundation were awarded in a grant to The University of Chicago which also managed the program in conjunction with National Astronomical Observatories, Chinese Academy of Sciences.



\bsp

\label{lastpage}


\begin{thebibliography}{99}

\bibitem[Aarseth(2003)]{aarseth2003} Aarseth, S.~J.\ 2003, 
Gravitational N-Body Simulations, Cambridge University Press, Cambridge, UK

\bibitem[Adams 
\& Laughlin(2001)]{adams2001} Adams, F.~C., \& Laughlin, G.\ 2001, \icarus, 150, 151 

\bibitem[Adams et al.(2006)]{adams2006} Adams, F.~C., Proszkow, 
E.~M., Fatuzzo, M., \& Myers, P.~C.\ 2006, \apj, 641, 504 

\bibitem[Adams(2010)]{adams2010} Adams, F.~C.\ 2010, \araa, 48, 47 

\bibitem[Allison et al.(2009)]{allison2009} Allison, R.~J., 
Goodwin, S.~P., Parker, R.~J., et al.\ 2009, \apjl, 700, L99 

\bibitem[Allison et al.(2010)]{allison2010} Allison, R.~J., 
Goodwin, S.~P., Parker, R.~J., Portegies Zwart, S.~F., 
\& de Grijs, R.\ 2010, \mnras, 407, 1098 

\bibitem[Bailey et al.(2014)]{bailey2014} Bailey, V., Meshkat, T., 
Reiter, M., et al.\ 2014, \apjl, 780, L4 

\bibitem[Batalha et al.(2013)]{batalha2013} Batalha, N.~M., Rowe, 
J.~F., Bryson, S.~T., et al.\ 2013, \apjs, 204, 24 

\bibitem[Bergfors et al.(2010)]{bergfors2010} Bergfors, C., Brandner, W., Janson, M., et al. \ 2010, \aap, 520, 54

\bibitem[Binney 
\& Tremaine(1987)]{binneytremaine} Binney, J., \& Tremaine, S.\ 1987, Princeton, NJ, Princeton University Press, 1987, 747 p.,  

\bibitem[Boley et al.(2012)]{boley2012} Boley, A.~C., Payne, 
M.~J., \& Ford, E.~B.\ 2012, \apj, 754, 57 

\bibitem[Bonnell et al.(2001)]{bonnell2001} Bonnell, I.~A., Smith, 
K.~W., Davies, M.~B., \& Horne, K.\ 2001, \mnras, 322, 859 

\bibitem[Borucki et al.(2011)]{borucki2011} Borucki, W.~J., Koch, 
D.~G., Basri, G., et al.\ 2011, \apj, 736, 19 

\bibitem[Boss(2011)]{boss2011} Boss, A.~P.\ 2011, \apj, 731, 74 

\bibitem[Bramich \& Horne(2006)]{bramich2006} Bramich D. M., Horne K., 2006, MNRAS, 367, 1677

\bibitem[Bressert et al.(2012)]{bressert2012} Bressert, E., Bastian, 
N., \& Gutermuth, R.\ 2012, Star Clusters in the Era of Large Surveys, 147 

\bibitem[Burke et al.(2006)]{burke2006} Burke, C.~J., Gaudi, 
B.~S., DePoy, D.~L., \& Pogge, R.~W.\ 2006, \aj, 132, 210 

\bibitem[Cai et al.(2015a)]{cai2015a} Cai, M.~X., Spurzem, R., 
\& Kouwenhoven, M.~B.~N.\ 2015a, arXiv:1501.01709 

\bibitem[Cai et al.(2015b)]{cai2015b} Cai, M.~X., Meiron, Y., Kouwenhoven, M.~B.~N., Assmann, P. \& Spurzem, R., \ 2015b, arXiv:1506.07591 

\bibitem[Cartwright 
\& Whitworth(2004)]{cartwright2004} Cartwright, A., \& Whitworth, A.~P.\ 2004, \mnras, 348, 589 

\bibitem[Cervi{\~n}o et 
al.(2013a)]{cervino2013a} Cervi{\~n}o, M., Rom{\'a}n-Z{\'u}{\~n}iga, C., Luridiana, V., et al.\ 2013, \aap, 553, AA31 

\bibitem[Cervi{\~n}o et 
al.(2013b)]{cervino2013b} Cervi{\~n}o, M., Rom{\'a}n-Z{\'u}{\~n}iga, C., Bayo, A., et al.\ 2013, \aap, 553, AA32 

\bibitem[Clarke et al.(2000)]{clarke2000} Clarke, C.~J., Bonnell, 
I.~A., \& Hillenbrand, L.~A.\ 2000, Protostars and Planets IV, 151 

\bibitem[Connelley et al.(2008a)]{connelley2008a} Connelley, M.~S., 
Reipurth, B., \& Tokunaga, A.~T.\ 2008a, \aj, 135, 2496 

\bibitem[Connelley et al.(2008b)]{connelley2008b} Connelley, M. S., Reipurth, B., \& Tokunaga, A. T. \ 2008b, \aj, 135, 2526

\bibitem[Craig 
\& Krumholz(2013)]{craig2013} Craig, J., \& Krumholz, M.~R.\ 2013, \apj, 769, 150 

\bibitem[Davies 
\& Sigurdsson(2001)]{davies2001} Davies, M.~B., \& Sigurdsson, S.\ 2001, \mnras, 324, 612 

\bibitem[Davies et al.(2013)]{davies2013} Davies, M.~B., Adams, 
F.~C., Armitage, P., et al.\ 2013, arXiv:1311.6816 

\bibitem[Delorme et al.(2012)]{delorme2012} Delorme, P., Gagn{\'e}, J., Malo, L., et al.\ 2012, \aap, 548, A26 

\bibitem[Di Stefano(2012)]{distefano2012} Di Stefano, R.\ 2012, 
\apjs, 201, 20 

\bibitem[Duquennoy 
\& Mayor(1991)]{duquennoy1991} Duquennoy, A., \& Mayor, M.\ 1991, \aap, 248, 485 

\bibitem[Dukes 
\& Krumholz(2012)]{dukes2012} Dukes, D., \& Krumholz, M.~R.\ 2012, \apj, 754, 56 

\bibitem[Fregeau et al.(2006)]{fregeau2006} Fregeau, J.~M., 
Chatterjee, S., \& Rasio, F.~A.\ 2006, \apj, 640, 1086 

\bibitem[Gaudi(2012)]{gaudi2012} Gaudi, B.~S.\ 2012, \araa, 50, 411 

\bibitem[Goodman 
\& Hut(1993)]{goodman1993} Goodman, J., \& Hut, P.\ 1993, \apj, 403, 271 

\bibitem[Goodwin 
\& Whitworth(2004)]{goodwin2004} Goodwin, S.~P., \& Whitworth, A.~P.\ 2004, \aap, 413, 929 

\bibitem[Goodwin 
\& Bastian(2006)]{goodwin2006} Goodwin S. P., Bastian N., 2006, MNRAS, 373, 752

\bibitem[Guenther et 
al.(2005)]{guenther2005} Guenther, E.~W., Paulson, D.~B., Cochran, W.~D., et al.\ 2005, \aap, 442, 1031 

\bibitem[Hao et al.(2013)]{hao2013} Hao, W., Kouwenhoven, 
M.~B.~N., \& Spurzem, R.\ 2013, \mnras, 433, 867 

\bibitem[Hartman et al.(2009)]{hartman2009} Hartman J. D. et al., 2009, ApJ, 695, 336

\bibitem[Heggie 
\& Hut(2003)]{heggiehut} Heggie, D., \& Hut, P.\ 2003, The Gravitational Million-Body Problem: A Multidisciplinary Approach to Star Cluster Dynamics, Cambridge University Press

\bibitem[Hurley \& Shara(2002)]{hurley2002} Hurley, J.~R., \& Shara, M.~M.\ 2002, \apj, 565, 1251 


\bibitem[Janson et al.(2014)]{janson2014} Janson, M., Bergfors, C., Brandner, W., et al. \ 2014, \apj, 789, 102

\bibitem[Jilkova et al.(2015)]{jilkova2015} Jilkova, L., Portegies 
Zwart, S., Pijloo, T., \& Hammer, M.\ 2015, arXiv:1506.03105 

\bibitem[Karttunen et al.(2003)]{karttunen2003} Karttunen, H., 
Kroeger, P., Oja, H., Poutanen, M., 
\& Donner, K.~J.\ 2003, Fundamental astronomy, by Hannu Karttunen, P.~Kroeger, H.~Oja, M.~Poutanen, and K.J.~Donner, 4th ed., Berlin: Springer, 2003.,  

\bibitem[Kouwenhoven et 
al.(2005)]{kouwenhoven2005} Kouwenhoven, M.~B.~N., Brown, A.~G.~A., Zinnecker, H., Kaper, L., \& Portegies Zwart, S.~F.\ 2005, \aap, 430, 137 

\bibitem[Kouwenhoven et 
al.(2007)]{kouwenhoven2007} Kouwenhoven, M.~B.~N., Brown, A.~G.~A., Portegies Zwart, S.~F., \& Kaper, L.\ 2007, \aap, 474, 77 

\bibitem[Kouwenhoven et al.(2010)]{kouwenhoven2010} Kouwenhoven, 
M.~B.~N., Goodwin, S.~P., Parker, R.~J., et al.\ 2010, \mnras, 404, 1835 

\bibitem[Kouwenhoven et al.(2014)]{kouwenhoven2014} Kouwenhoven, 
M.~B.~N., Goodwin, S.~P., de Grijs, R., Rose, M., 
\& Kim, S.~S.\ 2014, \mnras, 445, 2256 

\bibitem[Kraus et al.(2014)]{kraus2014} Kraus, A.~L., Ireland, 
M.~J., Cieza, L.~A., et al.\ 2014, \apj, 781, 20 

\bibitem[Kroupa(2001)]{kroupa2001} Kroupa, P.\ 2001, \mnras, 322, 
231 

\bibitem[Kruijssen(2012)]{kruijssen2012} Kruijssen, J.~M.~D.\ 2012, 
\mnras, 426, 3008 

\bibitem[K{\"u}pper et al.(2011)]{kupper2011} K{\"u}pper, 
A.~H.~W., Maschberger, T., Kroupa, P., 
\& Baumgardt, H.\ 2011, \mnras, 417, 2300 

\bibitem[Lada 
\& Lada(2003)]{lada2003} Lada, C.~J., \& Lada, E.~A.\ 2003, \araa, 41, 57 

\bibitem[Lafreni{\`e}re et al.(2008)]{lafreniere2008} Lafreni{\`e}re, 
D., Jayawardhana, R., \& van Kerkwijk, M.~H.\ 2008, \apjl, 689, L153 

\bibitem[Lagrange et 
al.(2009)]{lagrange2009} Lagrange, A.-M., Gratadour, D., Chauvin, G., et al.\ 2009, \aap, 493, L21 

\bibitem[Latham et al.(2011)]{latham2011} Latham, D.~W., Rowe, 
J.~F., Quinn, S.~N., et al.\ 2011, \apjl, 732, L24 

\bibitem[Laughlin 
\& Adams(1998)]{laughlin1998} Laughlin, G., \& Adams, F.~C.\ 1998, \apjl, 508, L171 

\bibitem[Li et al.(2015)]{li2015} Li, Y., Kouwenhoven, 
M.~B.~N., Stamatellos, D., \& Goodwin, S.~P.\ 2015, \apj, 805, 116 

\bibitem[Liu et al.(2013)]{liu2013} Liu, H.-G., Zhang, H., 
\& Zhou, J.-L.\ 2013, \apj, 772, 142 

\bibitem[Lucas et al.(2006)]{lucas2006} Lucas, P.~W., Weights, 
D.~J., Roche, P.~F., \& Riddick, F.~C.\ 2006, \mnras, 373, L60 

\bibitem[Malmberg et al.(2011)]{malmberg2011} Malmberg, D., Davies, 
M.~B., \& Heggie, D.~C.\ 2011, \mnras, 411, 859 

\bibitem[Marois et al.(2008)]{marois2008} Marois, C., Macintosh, 
B., Barman, T., et al.\ 2008, Science, 322, 1348 

\bibitem[Marois et al.(2010)]{marois2010} Marois, C., Zuckerman, 
B., Konopacky, Q.~M., Macintosh, B., \& Barman, T.\ 2010, \nat, 468, 1080 

\bibitem[Meibom et al.(2013)]{meibom2013} Meibom, S., Torres, G., 
Fressin, F., et al.\ 2013, \nat, 499, 55 

\bibitem[Mochejska at al.(2005)]{mochejska2005} Mochejska B.~J. et al., 2005, \aj, 129, 2856

\bibitem[Mochejska at al.(2006)]{mochejska2006} Mochejska B.~J. et al., 2006, \aj, 131, 1090

\bibitem[Moeckel 
\& Bate(2010)]{moeckel2010} Moeckel, N., \& Bate, M.~R.\ 2010, \mnras, 404, 721 

\bibitem[Moeckel 
\& Clarke(2011)]{moeckel2011} Moeckel, N., \& Clarke, C.~J.\ 2011, \mnras, 415, 1179 

\bibitem[Montalto at al.(2007)]{montalto2007} Montalto M. et al., 2007, \aap, 470, 1137

\bibitem[Nascimbeni et al.(2012)]{nascimbeni2012} Nascimbeni V., Bedin L. R., Piotto G., De Marchi F., Rich R. M., 2012, \aap, 541, A144

\bibitem[Naud et al.(2014)]{naud2014} Naud, M.-E., Artigau, 
{\'E}., Malo, L., et al.\ 2014, \apj, 787, 5 

\bibitem[Pacucci et al.(2013)]{pacucci2013} Pacucci, F., Ferrara, 
A., \& D'Onghia, E.\ 2013, \apjl, 778, L42 

\bibitem[Parker(2014)]{parkerdensity2014} Parker, R.~J.\ 2014, \mnras, 
445, 4037 

\bibitem[Parker 
\& Dale(2013)]{parkerdale2013} Parker, R.~J., \& Dale, J.~E.\ 2013, \mnras, 432, 986 

\bibitem[Parker et al.(2011)]{parker2011} Parker, R.~J., Goodwin, 
S.~P., \& Allison, R.~J.\ 2011, \mnras, 418, 2565 

\bibitem[Parker 
\& Quanz(2012)]{parkerquanz} Parker, R.~J., \& Quanz, S.~P.\ 2012, \mnras, 419, 2448 

\bibitem[Parker et al.(2014)]{parker2014} Parker, R.~J., Church, 
R.~P., Davies, M.~B., \& Meyer, M.~R.\ 2014, \mnras, 437, 946 

\bibitem[Parker et al.(2014)]{parkerwright2014} Parker, R.~J., Wright, 
N.~J., Goodwin, S.~P., \& Meyer, M.~R.\ 2014, \mnras, 438, 620 


\bibitem[Pelupessy et 
al.(2013)]{pelupessy2013} Pelupessy, F.~I., van Elteren, A., de Vries, N., et al.\ 2013, \aap, 557, AA84 

\bibitem[Perets 
\& Kouwenhoven(2012)]{perets2012} Perets, H.~B., \& Kouwenhoven, M.~B.~N.\ 2012, \apj, 750, 83 

\bibitem[Peretto et 
al.(2006)]{peretto2006} Peretto, N., Andr{\'e}, P., \& Belloche, A.\ 2006, \aap, 445, 979 

\bibitem[Pfalzner(2013)]{pfalzner2013} Pfalzner, S.\ 2013, \aap, 549, A82 

\bibitem[Portegies Zwart 
\& McMillan(2005)]{portegies2005} Portegies Zwart, S.~F., \& McMillan, S.~L.~W.\ 2005, \apjl, 633, L141 

\bibitem[Portegies Zwart(2009)]{portegies2009} Portegies Zwart, 
S.~F.\ 2009, \apjl, 696, L13 

\bibitem[Portegies Zwart et al.(2013)]{portegies2013} Portegies 
Zwart, S., McMillan, S.~L.~W., van Elteren, E., Pelupessy, I., 
\& de Vries, N.\ 2013, Computer Physics Communications, 183, 456 

\bibitem[Proszkow et al.(2009)]{proszkow2009} Proszkow, E.-M., 
Adams, F.~C., Hartmann, L.~W., \& Tobin, J.~J.\ 2009, \apj, 697, 1020 

\bibitem[Quanz et al.(2010)]{quanz2010} Quanz, S.~P., Goldman, 
B., Henning, T., et al.\ 2010, \apj, 708, 770 

\bibitem[Schmeja(2011)]{schmeja2011} Schmeja, S.\ 2011, 
Astronomische Nachrichten, 332, 172 

\bibitem[Shara et al.(2014)]{shara2014} Shara, M.~M., Hurley, 
J.~R., \& Mardling, R.~A.\ 2014, arXiv:1411.7061 

\bibitem[Smith 
\& Bonnell(2001)]{smith2001} Smith, K.~W., \& Bonnell, I.~A.\ 2001, \mnras, 322, L1 

\bibitem[Spitzer(1987)]{spitzer1987} Spitzer, L.\ 1987, Princeton, 
NJ, Princeton University Press, 1987, 191 p.,  

\bibitem[Spurzem et al.(2009)]{spurzem2009} Spurzem, R., Giersz, 
M., Heggie, D.~C., \& Lin, D.~N.~C.\ 2009, \apj, 697, 458 

\bibitem[Stamatellos 
\& Whitworth(2008)]{stamatellos2008} Stamatellos, D., \& Whitworth, A.~P.\ 2008, \aap, 480, 879 


\bibitem[Raghavan et al.(2010) ]{raghavan2010} Raghavan D.,McMaster H.A., Henry T. J., LathamD.W., Marcy G. W., Mason B. D., Gies D. R., White R. J., ten Brummelaar T. A.\ 2010, \apjs, 190, 1

\bibitem[Rosvick 
\& Robb(2006)]{rosvick2006} Rosvick J. M., Robb R. \ 2006, \aj, 132, 2309

\bibitem[Sumi et al.(2011)]{sumi2011} Sumi, T., Kamiya, K., 
Bennett, D.~P., et al.\ 2011, \nat, 473, 349 

\bibitem[Tokovinin(2014a)]{tokovinin2014a} Tokovinin, A.\ 2014a, \aj, 
147, 86 

\bibitem[Tokovinin(2014b)]{tokovinin2014b} Tokovinin, A.\ 2014b, \aj, 
147, 87 

\bibitem[Vorobyov(2013)]{vorobyov2013} Vorobyov, E.~I.\ 2013, \aap, 552, A129 

\bibitem[Wang et al.(2015a)]{wang2015a} Wang, L., Kouwenhoven, 
M.~B.~N., Zheng, X., Church, R.~P., 
\& Davies, M.~B.\ 2015a, \mnras, 449, 3543 

\bibitem[Wang et al.(2015b)]{wang2015b} Wang, L., Spurzem, R., 
Aarseth, S., et al.\ 2015b, \mnras, 450, 4070 

\bibitem[Weidner 
\& Kroupa(2004)]{weidner2004} Weidner, C., \& Kroupa, P.\ 2004, \mnras, 348, 187 

\bibitem[Weidner et al.(2013)]{weidner2013} Weidner, C., Kroupa, 
P., \& Pflamm-Altenburg, J.\ 2013, \mnras, 434, 84 

\bibitem[Weldrake et al.(2005)]{weldrake2005} Weldrake D. T. F., Sackett P. D., Bridges T. J., Freeman K. C., 2005, ApJ, 620, 1043

\bibitem[Weldrake et al.(2008)]{weldrake2008} Weldrake D. T. F., Sackett P. D., Bridges T. J., 2008, ApJ, 674, 1117

\end{thebibliography}
\end{document}